\documentclass[preprint,showpacs,preprintnumbers,amsmath,amssymb]{revtex4}

\usepackage{graphicx}\graphicspath{./graph/}%
\usepackage{dcolumn}% Align table columns on decimal point
\usepackage{bm}% bold math
\usepackage{color}
\usepackage{verbatim}
\usepackage{amssymb}
\usepackage{amsfonts}
\usepackage{latexsym}
\usepackage{epstopdf}
\usepackage{url}
\definecolor{red}{rgb}{1,0,0}
\definecolor{green}{rgb}{0,1,0}
\definecolor{blue}{rgb}{0,0,1}

\newcommand{\be}{\begin{equation}}
\newcommand{\ee}{\end{equation}}
\newcommand{\bea}{\begin{eqnarray}}
\newcommand{\eea}{\end{eqnarray}}
\newcommand{\bdm}{\begin{displaymath}}
\newcommand{\edm}{\end{displaymath}}

\newcommand\ptl{\partial}
\newcommand\dpr{^{\prime\prime}}

\newcommand\pr{^\prime}

\newcommand\imp{{\rm Im}}
\newcommand\rep{{\rm Re}}

\begin{document}
%\preprint{APS/123-QED}
\title{Equilibrium of an Arbitrary Bunch Train\\ in Presence of a Passive Harmonic Cavity:\\
Solution through Coupled Ha\" issinski Equations }
\author{Robert Warnock }
\email{warnock@slac.stanford.edu}
\affiliation{SLAC National Accelerator Laboratory, Stanford University, Menlo Park, CA 94025, USA}
\affiliation{Department of Mathematics and Statistics, University of New Mexico, Albuquerque, NM 87131, USA}
\author{Marco Venturini}
\email{mventurini@lbl.gov}
\affiliation{Lawrence Berkeley National Laboratory,  University of California, Berkeley, CA 94720, USA}

\begin{abstract}
We study the effect of a passive harmonic cavity, introduced to cause bunch lengthening, in an electron storage ring.
We derive a formula for the induced voltage from such a cavity with high $Q$, excited by a
a sequence of bunches, allowing for arbitrary gaps in the sequence and arbitrary currents. Except for a minor term that can be determined iteratively, the voltage is given in terms of a single mode of the Fourier transforms of the bunch forms, namely the mode at the resonant frequency of the cavity.  Supposing that the only wake field is from the harmonic cavity, we derive a system of coupled Ha\"issinski equations which determine
the bunch positions and profiles in the equilibrium state. The number of unknowns in the system is only twice the number of bunches, and it can be solved quickly by a Newton iteration, starting with a guess determined
by path-following from a solution at weak current.  We explore the effect of the fill pattern
on the bunch lengthening, and also the dependence on the shunt impedance and detuning of the cavity away from the third harmonic of the main accelerating cavity. We consider two measures to reduce the effects of gaps: 1)~ distribution of the gaps around the ring to the greatest extent allowed, and 2)~`` guard bunches" with higher charges adjacent to the gaps, compensating for the charge missing in gaps.    Results for parameters of the forthcoming ALS-U light source are
presented.
\end{abstract}
\maketitle
\section{Introduction}
In electron storage rings the phenomenon of Touschek scattering often limits the lifetime of a stored beam \cite{piwinski}. This is the aspect of intrabeam scattering in which small transverse momenta are transformed through Coulomb scattering and a Lorentz boost into one large and one small longitudinal momentum in the lab frame,  sending both particles outside the momentum aperture of the ring. The effect may be counteracted by reducing the charge density in the beam. One way to do that is to increase the bunch size in the longitudinal direction. This can be done by adding a cavity with resonant frequency close to a low harmonic of the main r.f. frequency, say the third or fourth harmonic. This is often called
an HHC (higher harmonic cavity).

The quadratic potential well of a usual r.f. system can be turned into a quartic potential well,
by arranging the HHC so as to zero the second and third derivatives of the effective well \cite{hofmann,byrd1}.  This
condition, often referred to as  ``ideal", results in a flat-top equilibrium bunch profile
with substantial bunch length increase, say by a factor of four or more in cases of interest,
and an increase of the Touschek lifetime by a comparable factor. However, the flat top is not necessarily the best configuration, since a further lifetime improvement can be achieved
by ``over-stretching", which causes the appearance of two peaks in the bunch profile.
This must not  be carried too far, however, since eventually the average lifetime will degrade rather than improve with over-stretching.

Higher-harmonic cavities  for bunch lengthening  have long been in use at several
3rd generation light sources, including MAX-II \cite{georgsson}, ALS \cite{byrd_comm}, SLS \cite{pedrozzi}, ELLETRA \cite{svandrlik},
BESSY \cite{anders}) and in the DAFNE $e_+e_-$ collider \cite{alesini}. There is  now renewed interest
driven by the trend toward ultra-low emittance, which is making the new generation of  light sources increasingly
sensitive to scattering effects.  A possible
installation at NSLS-II is being evaluated \cite{nsls2,blednych,bassi}, and the 4th generation light source
MAX-IV has a system already commissioned \cite{skripka}. Forthcoming 4th generation machines,
including ALS-U \cite{feng}, APS-U \cite{kelly}, SIRIUS \cite{sirius}, PETRA-IV \cite{schroer},
 SLS-2 \cite{streun},  ESRF-EBS \cite{esrf},  Diamond-II  \cite{diamondII}, and HLS-II \cite{hlsII}
 all have harmonic cavities as essential components.

The harmonic cavity may be passive or actively excited, but a natural first step is to consider the
less expensive passive option. Our discussion is for the passive case, but our methods could be adapted
to the active system. In the passive case, the field induced in the cavity by a bunch train depends
strongly on the fill pattern. If the beam has a uniform fill pattern, {\sl e.g.}
all r.f. buckets are filled or all the bunches are separated by a fixed number of empty r.f. buckets,
there exists a beam equilibrium with all bunches having the same profile (possibly
of the flattop form if the ideal HHC settings are met). However,  if there are significant
gaps between bunch trains (or a long gap following a single-train beam)  the quality of the beam
equilibrium  can be compromised. Instead of uniform charge distributions along the train, one then
sees a variation  of the bunch form and centroid along the train. This
may cause  severe limitations to the effectiveness of the HHC system,  either because of the resulting  uneven lifetime
or/and because of interference with the functioning of the machine feedback systems
used for beam stabilization, and may prevent the attainment of the desired  bunch lengthening.

There are several reasons for the presence of gaps in the bunch train. Historically, gaps have been needed for ion clearing. Another
demand arises from the requirements of synchrotron light users, who may need different fill patterns for different types
of experiments. Experiments needing precise timing of x-ray pulses generally require more gaps than those asking for high brilliance.
In the ALS-U, gaps are needed for on-axis injection from the accumulator ring \cite{steier}.

In this paper we present a robust and efficient method to evaluate the beam equilibrium for arbitrary HHC
settings and beam-fill patterns. Our approach, entailing the numerical solution of a system of non-linear algebraic
equations, extends the method introduced in \cite{bobkarl}  for the
determination of single-bunch Ha\"issinski equilibria in the case of short-range wake fields.  It is much faster than macro-particle based methods and,
we believe, an improvement on the method recently introduced by T. Olsson {\sl et al.}, \cite{olsson}.

Our immediate objective  is to study the effect of the fill pattern on the bunch densities
in the equilibrium state. While this is a useful first step with rewarding  practical implications, {\sl e.g.}
offering guidance on  choice of the HHC design parameters, our final goal is to understand
the threshold in current for an instability, and the time-dependent behavior beyond the threshold.

 The widened potential well has
some benefits: the reduced peak bunch current and increased longitudinal tune spread
may lead to the damping of certain instabilities. However, other instabilities may be induced,
 either through the fundamental or higher order modes of the HHC \cite{krinsky, bosch, marco1, marco2},
 or by possibly aggravating the effect of higher order modes in the main cavity  \cite{ryan}.
The method presented here is an essential ingredient toward the application of
mode-analysis techniques to the study of beam stability when HHC's are present.

Beside reports on specific projects as cited above, there are several papers  which discuss the issues that concern us
in a more or less general way, through theory, simulations, and  measurements. Byrd and Georgsson \cite{byrd1} and Hofmann and Myers \cite{hofmann} treated the situation without HHC beam loading (i.e., without the cavity wake field), which is the starting point for the present work. Towne \cite{towne} studied
stability of stretched bunches in the presence of  a broad band impedance together with a high-$Q$ resonator, using Vlasov-Fokker-Planck simulations and measurements at NSLS-VUV. Byrd, De  Santis, Jacob and Serriere \cite{byrd2} initiated the study of the impact of gaps in
the bunch train. They used the term ``transient beam loading", which several authors have adopted. (Since a transient effect is usually
thought of as short-lived in {\it time}, not the case here, ``inhomogeneous beam loading" might be a more descriptive term.) A direct antecedent
of our work is the paper of Tavares, Andersson, Hansson, and Breunlin \cite{tavares} who were concerned with self consistency in the equilibrium bunch densities. The study of this topic was continued by Olsson, Cullinan, and Andersson \cite{olsson} who developed an iterative scheme to find the equilibrium charge densities.  Bassi and Tagger \cite{bassi} investigated the option of a super-conducting HHC, invoking self-consistent simulations and emphasizing the importance of beam loading in the main accelerating cavity for a full picture.

The content of the paper is as follows:

 Section \ref{section:variables} describes our choice of coordinates and the description of the bunch train. Section \ref{section:eom} and Appendix \ref{section:appA} review the equations of motion.

 Section \ref{section:primary} states the primary
formula for the voltage induced by the harmonic cavity, then  Section \ref{section:effective} notes that the induced voltage can be expressed in terms of an effective wake potential, which is represented by a compact formula that is the basis for further work. Section \ref{section:induced} goes on to find an explicit formula for the induced voltage from
an arbitrary bunch train, which is in terms of the Fourier transforms of the bunch forms at the resonant frequency of the harmonic cavity.

 Section \ref{section:vfp} states  the Vlasov-Fokker-Planck equation, and shows how its steady state solution is given by the solution of coupled Ha\"issinski equations. Section \ref{section:mean_E_trans} shows that the mean energy transfer in the equilibrium state is exactly equal to the energy loss per turn. Section \ref{section:integral} calculates the integral of the induced voltage, to get the potential wells for the Ha\"issinski system.

  Section \ref{section:newton} describes a Newton iteration for solution of the Ha\"issinski system, while
Section \ref{section:jacobian} gives the associated Jacobian matrix, and Section \ref{section:current} shows how to follow the Newton solution
as a function of current.

Section \ref{section:numerical} presents numerical results for the parameters of ALS-U and a comparison to a macro-particle simulation.  Section \ref{section:touschek} estimates Touschek lifetimes as a function of the cavity detuning.

Appendix \ref{section:appB} discusses the perturbation of the synchronous phase due to the harmonic cavity, and reports that there
is no necessity to base the coordinate system on the perturbed phase. Appendix \ref{section:appC} explains how our general formula for the induced voltage reduces to a known formula in the case where all bunches are identical.

\section{Choice of variables and description of bunch train \label{section:variables}}
Synchrotron motion in a storage ring can be described in terms of
the longitudinal coordinate $z=\beta_0 ct-s$, the distance to the reference particle. Here $s$ measures position
in the laboratory
as  arc length along a reference trajectory, and the reference particle  has position  $s_0=\beta_0 ct$
at time $t$.  Particles leading the reference particle have $z<0$. The opposite sign convention is often adopted, indeed
in our own papers.

 For a single bunch, $z$ is familiar as the
``beam frame coordinate", which is suitable as a phase space coordinate for equations of motion and the Vlasov equation. In the case of many bunches, $z$ is a convenient global coordinate for description of the total charge density, and merely by adding
constants to $z$ we can construct local beam frame coordinates for all the bunches. Moreover, $z$ has the convenient
property of being proportional to $s$ at fixed $t$ and proportional to $t$ at fixed $s$.
 Thus if we wish to demonstrate periodicity in $s$ at a fixed time we have
only to demonstrate periodicity in $z$.

We consider a sequence of $n_b$ bunches , giving a total charge density of the form
\be
\rho_{\rm tot}(z)=\sum_{p=-\infty}^\infty~\sum_{j=1}^{n_b} ~\xi_j\rho_j(z+m_j\lambda_1+pC) , \label{train}
\ee
where $\lambda_1$ is the wavelength of the main r.f. cavity, and $C$ is the circumference of the ring. The $m_j$
are non-negative integers specifying the filled r.f. buckets. Without loss of generality we take $m_1=0$; then
$m_j\le h-1$, where $h$ is the harmonic number, equal to the maximum number of bunches, and $h\lambda_1=C$. We take $\int\rho_j(z)dz=1$, and define $\xi_j$ as the ratio of the charge in bunch $j$ to the average  bunch charge. The  leading bunch in a train, having the most negative $z$, has the highest bunch index, $j=n_b$.

The bunch profiles $\rho_j(z)$ are time-independent, since we are concerned with the equilibrium state, and are initially unknown functions to be determined by the condition of equilibrium.

The infinite sequence in (\ref{train}) is intended to mimic the periodicity of the charge density in
a circular storage ring. We have $\rho_{\rm tot}(z+C)=\rho_{\rm tot}(z)$, so that at fixed $t$ the density is periodic in $s$ with period $C$. At fixed $s$ it is also periodic in $t$ with period $C/\beta c$.
 The idealization of supposing that the charge pattern exists for all $t\in (-\infty,\infty)$  is justified, given the  large storage times of typical machines.

The total voltage seen by a particle at arbitrary $z$ (at an arbitrary distance from the reference particle)  is taken to be
\be
V_1\sin(k_1z+\phi_0)+V_r(z)\ , \label{vtot}
\ee
where $k_1=2\pi/\lambda_1$. In the model to be explored, the induced voltage $V_r$ comes only from the lowest mode of the passive harmonic cavity, as excited by the bunch train. The relation of $\phi_0$ to the synchronous phase, the phase at which the cavity supplies the mean energy lost per turn, will be discussed presently.

 We define $z_j$,  the argument of the density $\rho_j$, as
\be
z_j=z+m_j\lambda_1\ .  \label{zj}
\ee
 Then by (\ref{vtot})
the total voltage as a function of $z_j$ is
\be
V_1\sin(k_1z_j+\phi_0)+V_r(z_j-m_j\lambda_1)\ ,      \label{vj}
\ee
since the first term in (\ref{vtot}) is periodic in $z$ with period $\lambda_1$.

\section{Equations of motion \label{section:eom}}
The usual equations of motion for a single particle, subject only to  applied r.f.,  describe oscillations in
a potential well with minimum at the location of the synchronous particle. Since the harmonic cavity broadens the well and shifts its minimum, a natural step would be to modify the equation of motion  so that it describes oscillations
about the shifted minimum. On the other hand, this might be an unnecessary complication
if the shift is sufficiently small. The coordinate of the unperturbed problem might provide a
perfectly accurate description, even if it is not the distance to the minimum.

  We first recall the derivation of the standard equations for a single particle with only applied r.f. We first derive difference equations, referring to
to changes over a full turn, and later replace them by differential equations, since the changes are very small. The salient variable of interest is the phase $\phi$ of the applied r.f.
at the time that the particle crosses the accelerating cavity.

At the $n$-th turn the r.f. kick at phase $\phi_n$ restores the energy loss $U_0$ of the previous turn and also changes the energy of a generic particle from $E_n$ to $E_{n+1}$:
\be
(U_0+E_{n+1})-E_n=eV_1\sin\phi_n\ .     \label{dE}
\ee
The synchronous phase $\phi_0$ is that for which the energy supplied is exactly $U_0$:
\be
eV_1\sin\phi_0=U_0\ . \label{phis0}
\ee
For stable motion this angle should be in the second quadrant,
\be
\cos\phi_0=-(1-\sin^2\phi_0)^{1/2}\ ,  \label{stab}
\ee
with the  square root defined to be positive.
Defining
\be
(\Delta E)_n=E_n-E_0\ ,\qquad (\Delta\phi)_n=\phi_n-\phi_0\ ,  \label{deldefs}
\ee
where $E_0$ is the nominal energy of the ring, we write (\ref{dE}) as
\be
 (\Delta E)_{n+1}-(\Delta E)_n=eV_1\sin((\Delta\phi)_n+\phi_0)-U_0\ . \label{firstdiff}
\ee

The change with $n$ of $(\Delta\phi)_n$ depends on the revolution frequency, which in turn depends on $(\Delta E)_n$, these dependencies being linear to a good approximation. Invoking the definition of the momentum compaction factor $\alpha$
we show in Appendix \ref{section:appA} that
\be
(\Delta\phi)_{n+1}-(\Delta\phi)_n=\alpha k_1C\frac{(\Delta E)_n}{E_0}\ .   \label{seconddiff}
\ee

We wish to follow the trajectory of $z$, namely
\be
z(t)=\beta_0ct-s(t)\ , \label{ztraj}
\ee
which is related to the trajectory of $\phi$ as follows:
\be
\frac{\lambda_1}{2\pi}(\phi(t)-\phi_0)={\rm signed~ distance~ to~ the~ reference~ particle} = z(t)\ , \label{dphiz}
\ee
hence
\be
 (\Delta\phi)_n=k_1z_n\ . \label{dphin}
\ee
The sign in (\ref{dphiz}) is correct: if $\phi(t) >\phi_0$ at time $t$ when the particle arrives at the cavity, it has arrived later
than the reference particle, which is to say that $s(t)<\beta_0ct$.

Approximating the difference equations (\ref{firstdiff}) and (\ref{seconddiff}) by
differential equations, with $dt=T_0$ and $\delta=(E-E_0)/E_0$, and applying (\ref{dphin}) we have
\bea
 &&\frac{d\delta}{dt}=\frac{1}{E_0T_0}\big(eV_1\sin(k_1z+\phi_0)-U_0\big)\ ,\label{firstde}\\
 &&\frac{dz}{dt}=\alpha \beta_0c\delta\ .                       \label{secondde}
\eea
Here $T_0=C/\beta_0c$ is the nominal revolution time of the ring. In replacing (\ref{firstdiff}) by (\ref{firstde}) we have equated $T_0$ with the time between successive arrivals at the cavity, but this is correct at best  in an average sense, because different particles have different revolution times. This approximation is not usually acknowledged in textbook treatments of the problem.

The generalization of (\ref{firstde}) and (\ref{secondde}) to account for many bunches and the harmonic cavity is
obtained by invoking the total voltage (\ref{vtot}) and the replacements $z\rightarrow z_i-m_i\lambda_1\ ,\quad \delta \rightarrow \delta_i$, thus
\bea
 &&\frac{d\delta_i}{dt}=\frac{1}{E_0T_0}\big(eV_1\sin(k_1z_i+\phi_0)+eV_r(z_i-m_i\lambda_1)-U_0\big)\ ,\label{firstdenb}\\
 &&\frac{dz_i}{dt}=\alpha \beta_0c\delta_i\ , \hskip 3cm i=1,\cdots,n_b\ .                      \label{seconddenb}
\eea
For this we note that the required relation (\ref{seconddiff}) is derived in Appendix \ref{section:appA} with allowance for
the presence of $V_r$. The derivation requires $V_r(z)=V_r(z+C)$, which is assured in the following
formalism.

A new feature is that $\phi_0$ is no longer the synchronous phase, since the induced voltage $V_r$ causes an
additional energy increment that must be taken into account.  We are nevertheless free to choose $\phi_0$ according to (\ref{phis0}) and (\ref{stab}), and we shall indeed make that choice. Some nearby value could do as well.

In order to clarify the impact of the shifted synchronous phase, we have also carried out a calculation with the coordinate system shifted accordingly. We conclude that there is no need to work in such a system. This issue is reviewed in Appendix \ref{section:appB}.

\section{Primary formula for the induced voltage } \label{section:primary}
At an arbitrary $z$ the induced voltage from the harmonic cavity will be
\bea
&&V_r(z)=-eN\int_{-\infty}^\infty W(z-z\pr)\rho_{\rm tot}(z\pr)dz\pr\nonumber\\&&=-eN\int_{-\infty}^\infty W(z-z\pr)
\sum_{p=-\infty}^\infty~\sum_{j=1}^{n_b} ~\xi_j\rho_j(z\pr+m_j\lambda_1+pC)dz\pr\ . \label{induced}
\eea
Here $W$ is the wake potential of the cavity, which for sufficiently large $Q$ has the form
\be
W(z)=\frac{\omega_rR_s}{Q}\theta(z)e^{-k_rz/2Q}\cos(k_rz)\ . \label{W}
\ee
In this formula $\omega_r=k_r c$ is the circular resonant frequency of the lowest mode of the cavity, $R_s$ is its shunt impedance,
$Q$ its quality factor, and $\theta(z)$ is the unit step function, equal to 1 for $z\ge 0$ and 0 otherwise. The $\theta$ function is an expression of causality.

The expression (\ref{induced}) satisfies the obvious requirement that $V_r$ be periodic with period $C$. To see that,
evaluate $V_r(z+C)$ by changing the integration variable to $z\dpr=z\pr-C$ and the summation variable to $p\pr=p+1$.

We suppose that the support of any $\rho_j(z)$, the region in which it is non-zero, is much less in extent than $\lambda_1$, a
condition that is satisfied in any ring of interest for this study.

To proceed it is convenient to translate the variable of integration and reverse the order of integration and summation, so that the formula (\ref{induced}) takes the form
\be
V_r(z)=-eN\int_{-\infty}^\infty~\sum_{p=-\infty}^\infty W(z-z\pr+pC)~\sum_{j=1}^{n_b} \xi_j\rho_j(z\pr+m_j\lambda_1)dz\pr\ .
\label{primary}
\ee

To give an idea of typical parameters for the following work, we list in Table I  a tentative set of parameters for ALS-U, the forthcoming upgrade of the Advanced Light Source at Lawrence Berkeley National Laboratory.
\begin{table}
\begin{center}
\caption{~~Baseline parameters for ALS-U design}
\vskip .2cm
%\begin{ruledtabular}
\setlength{\tabcolsep}{.25in}
\begin{tabular} {l|c|c}
Ring circumference& $C$&196.5 ~m\\
Beam energy&$E_0$&2~ GeV\\
Average bunch current&$I_{\rm avg}$&500~mA\\
Momentum compaction&$\alpha$&$2.11\times10^{-4}$\\
Natural energy spread&$\sigma_\delta$&$9.43\times10^{-4}$\\
Natural rms bunch length&$\sigma_{z0}$&3.54 mm\\
Energy loss per turn&$U_0$&0.217 MeV\\
Harmonic number&$h$&328\\
Main cavity frequency&$f_1$&500.417 MHz\\
Main cavity voltage&$V_1$&0.6 MV\\
Harmonic cavity harmonic number &$3$&\\
Harmonic cavity frequency&$f_r$&1501.84 MHz\\
Harmonic cavity detuning&$\delta f=f_r-3f_1$&0.2502 MHz\\
Harmonic cavity shunt impedance&$R_s$&1.39 M$\Omega$\\
Harmonic cavity  quality factor&$Q$&$2\times10^4$\\
\end{tabular}
\end{center}
\label{table: table1}
\end{table}

\section{Effective wake potential } \label{section:effective}

The sum over $p$ in (\ref{primary}) can be thought of as an effective wake potential $\mathcal{W}(z)$
which is to be convolved with an effective  charge density $\rho(z)$, defined as follows:
\be
\mathcal{W}(z)=\sum_{p=-\infty}^\infty W(z+pC)\ ,\quad\rho(z)=\sum_{j=1}^{n_b}\xi_j\rho_j(z+m_j\lambda_1)\ .\label{wrho}
\ee
Then the induced voltage may be expressed as
\be
V_r(z)= -eN\int_{-\infty}^\infty \mathcal{W}(z-z\pr)\rho(z\pr)dz\pr\ .   \label{indv}
\ee

Applying (\ref{W}) and expanding the cosine by the double angle formula we have
\be
\mathcal{W}(z)=\frac{\omega_rR_s}{Q}\sum_{p=-\infty}^\infty\theta(z+pC)e^{-k_r(z+pC)/2Q}
\big[\cos(k_rz)\cos(pk_rC)-\sin(k_rz)\sin(pk_rC)\big]\ .\label{weffarb}
\ee
The $\theta$ function requires $p\ge -z/C$, but since $p$ is an integer that means
\be
p\ge p_0=\lceil-z/C\rceil\ ,    \label{ceil1}
\ee
where $\lceil x \rceil$ denotes the ceiling of $x$, which is the smallest integer greater than or
equal to $x$.
Expressing the sine and cosine of $pk_rC$ in terms of exponentials, we find that
\be
\mathcal{W}(z)
=\frac{\omega_rR_s}{Q}e^{-k_rz/2Q}\big[\cos(k_rz)\rep\zeta(k_r,z)+\sin(k_rz)\imp\zeta(k_r,z)\big]\ ,\label{wzeta}
\ee
where
\be
\zeta(k_r,z)=\sum_{p=p_0}^\infty r^{ph}=r^{p_0h}\sum_{p=0}^\infty r^{ph}=r^{p_0h}\frac{1}{1-r^h}\ ,\label{zeta}
\ee
with
\be
r=\exp\big[-k_r\lambda_1(i+1/2Q)\big] .\label{rdef}
\ee

It is convenient to define real polar variables $(\eta(k_r),\psi(k_r))$ such that
\be
\frac{1}{1-r^h}=\eta e^{-i\psi}\ .   \label{polar}
\ee
Then  from (\ref{zeta}), (\ref{rdef}) and (\ref{polar}) we have
\be
 \rep\zeta=\eta e^{-p_0k_rC/2Q}\cos(p_0k_rC+\psi)\ ,\quad \imp\zeta=-\eta e^{-p_0k_rC/2Q}\sin(p_0k_rC+\psi)\ .
 \label{reimzeta}
\ee
\begin{figure}[htb]
   \centering
   \includegraphics[width=0.7\linewidth]{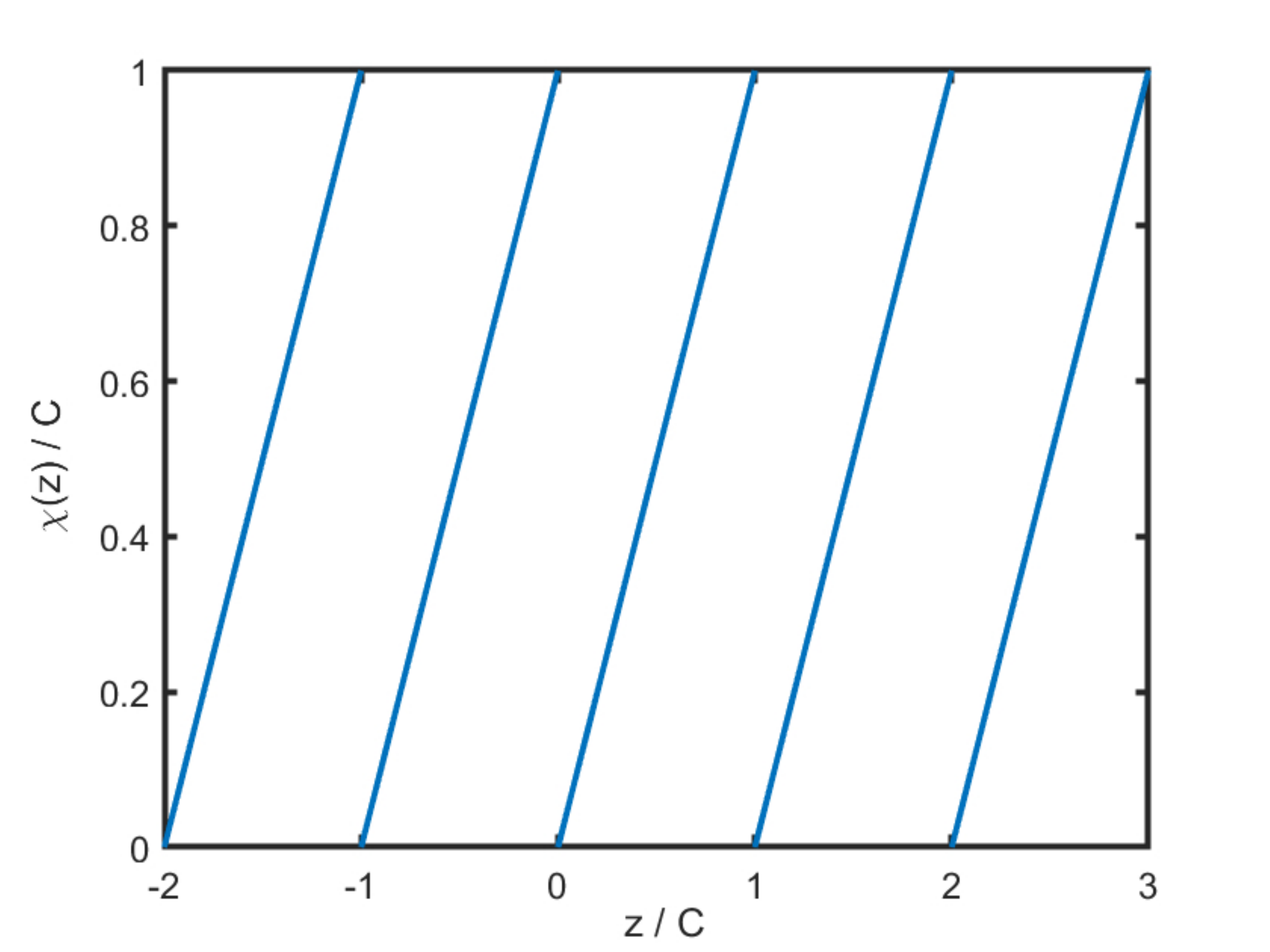}
  \caption{The sawtooth function $\chi(z)$. Its value at jumps is defined by the limit from the left.}
   \label{fig:fig1}
\end{figure}
Substituting in (\ref{wzeta}) and applying the double angle formula in reverse we have
\be
\mathcal{W}(z)=\frac{\omega_rR_s}{Q}\eta(k_r)\exp(-k_r\chi(z)/2Q)\cos\big(k_r\chi(z)+\psi(k_r)\big)\ ,
   \label{weff}
 \ee
 where
 \be
\chi(z)=z+C\lceil -z/C\rceil= z+nC \quad {\rm for}\quad nC < z \le (n+1)C\ ,\quad n={\rm integer}\ . \label{xi}
\ee
The function $\chi(z)$, plotted in Fig.\ref{fig:fig1}, is periodic with period $C$ and has a sawtooth form, with its value $C$ at jumps defined by the limit from the left. It follows that $\mathcal{W}(z)$ and the induced voltage $V_r(z)$ defined by (\ref{indv}) are periodic with period $C$.

In (\ref{weff}) we have an appealing, compact formula for the effective wake potential, which will lead to
the induced voltage after a straightforward evaluation of the integral (\ref{indv}). One must keep in mind that
the integrand in (\ref{indv}) has a jump at $z=z\pr$ owing to the jump in $\chi(z)$ at $z=0$.

\begin{figure}[htb]
   \centering
   \includegraphics[width=0.8\linewidth]{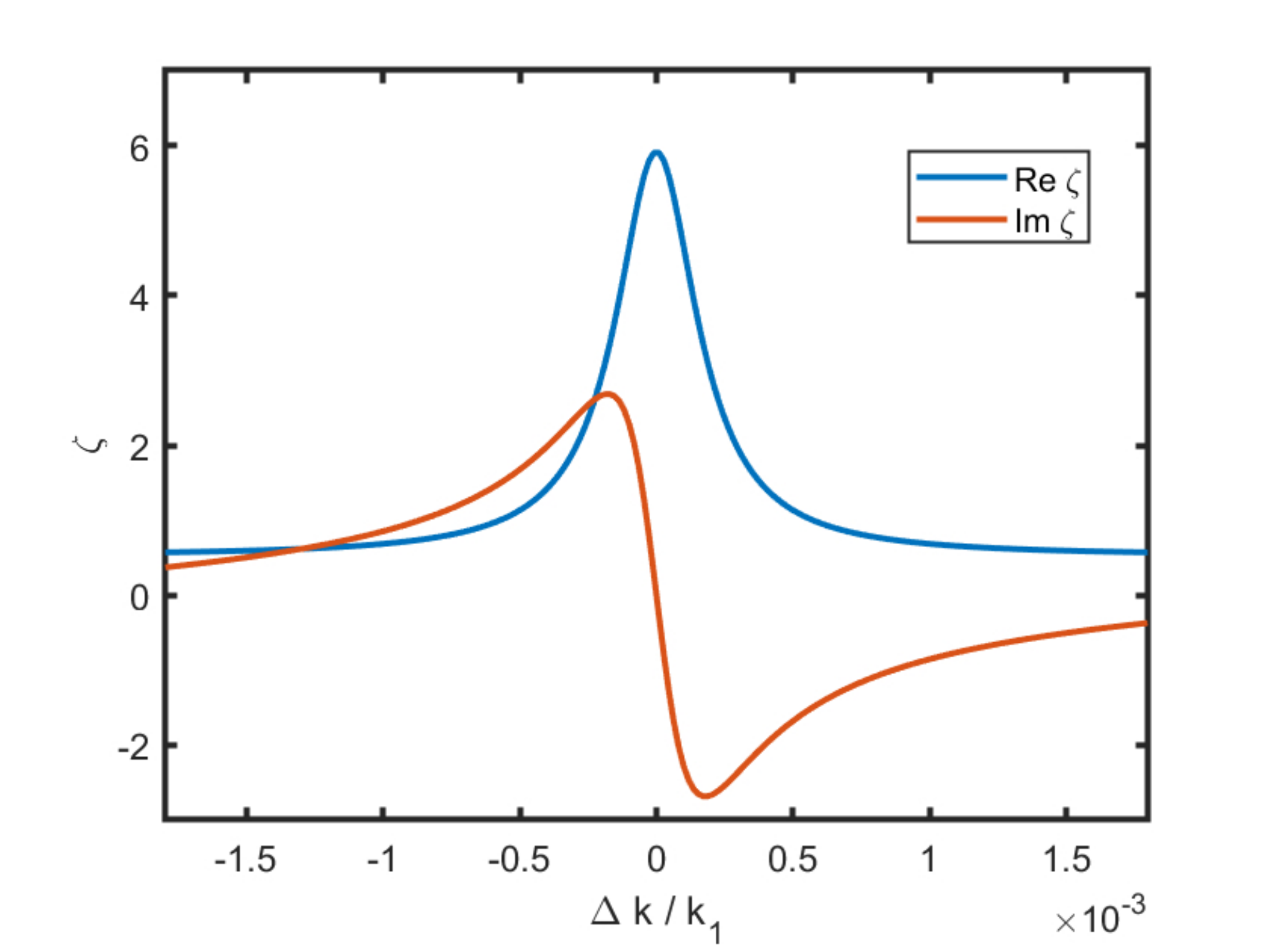}
  \caption{ $\zeta=\eta\exp(-i\psi)$ as a function of the detuning parameter $\Delta k/k_1$. The half-width of the peak in
 the real part is roughly $3/(2Q)=0.4\cdot 10^{-4}$.}
   \label{fig:fig2}
\end{figure}

An important feature of $\mathcal{W}(z)$ is its  behavior as a function of the detuning parameter $\Delta k/k_1$, where for a 3rd harmonic cavity $\Delta k=k_r-3k_1$. Figure \ref{fig:fig2} shows the real and imaginary parts of the function $\zeta=\eta\exp(-i\psi)$ for a typical case.
The function resembles a Lorentzian resonant line form, but in fact differs substantially from an actual Lorentzian. The half-width of the peak in
the real part is roughly $3/(2Q)$. Later we shall find that a true Lorentzian with that half-width occurs in the case of a complete fill with $h$ identical bunches. Since one can show that $\zeta$ approaches a true Lorentzian as $\Delta k$ tends to zero, the maximum of $\eta$ is exactly at
$\Delta k=0$.

\section{The induced voltage \label{section:induced}}
We are now in a position to compute the induced voltage from (\ref{indv}).
The density $\rho$, defined in (\ref{wrho}), is zero except for $n_b$ isolated peaks, the bunch profiles.
We define the interval $\Omega_i$ which is to contain the support of the $i$-th bunch, much shorter than the
main r.f. wavelength:
\be
\Omega_i=\big\{z\big|-\Sigma\le z+m_i\lambda_1 \le \Sigma \big\}\ ,\qquad  2\Sigma~\ll~\lambda_1\ .          \label{sigj}
\ee
This is just to say that the support in terms of the beam frame coordinate $z_i$ is within the region $| z_i|\le \Sigma$.
Note that the elements of $\Omega_i$ are close to $z=-m_i\lambda_1$ and therefore decrease with increasing $i$.
 Because of the stated restriction on $\Sigma$, no  two of the $\Omega_i$ can intersect:
\be
\Omega_i\cap\Omega_j=\emptyset\ , \quad i\ne j.
\ee
Some numerical experimentation may be needed to find an appropriate and economical value of $\Sigma$.

We note  that $V_r(z)$ need be evaluated only at $z$ within the various $\Omega_i$, since the collective force enters the dynamics only in those regions, through the Ha\"issinski or Vlasov equations. Also, $\rho(z\pr)$ is non-zero only for $z\pr$ in the same sets. It follows that the
function $\chi(z-z\pr)$ in (\ref{weff}) takes on only two values:
\be
\chi(z-z\pr)=\left\{\begin{array} {cc} z-z\pr\ , &\quad z\pr <z\\ z-z\pr +C\ , &\quad z\pr > z\\ \end{array}\right.\ ,\quad
z\in\Omega_i\ ,\ z\pr\in\Omega_j\ . \label{chivalues}
\ee
This follows from the fact that $|z-z\pr|<C$. Regardless of the fill pattern, $|z-z\pr|$ cannot be greater than a
number  $C-\lambda_1+\mathcal{O}(\Sigma)$. (For instance, if we have only two buckets, $C=2\lambda_1$, then the distance between a particle in one bucket and a particle in the other is $\lambda_1$ plus a quantity of order $\Sigma$ .) Thus
\be
\lceil -(z-z\pr)/C\rceil=\left\{\begin{array} {cc} 0\ , &\quad z\pr <z\\ 1\ , &\quad z\pr > z\\ \end{array}\right.\ ,
\label{ceilvalues}
\ee
which implies (\ref{chivalues}).

Let us evaluate $V_r(z)$
for $z\in\Omega_i$, and divide the terms into three groups, those for which $z<z\pr$, those for which $z>z\pr$, and the
one ``diagonal" term in which both $z<z\pr$ and $z>z\pr$ can occur. Thus
\be
V_r(z)=\frac{-eN\omega_rR_s}{Q}\eta(k_r)(v^<(z)+v^>(z)+v^d(z))\ ,\quad z\in\Omega_i\ , \label{decmp}
\ee
where
\bea
&&v^<(z)=\sum_{j=1}^{i-1}\xi_j\int_{\Omega_j}
\exp(-k_r(z-z\pr+C)/2Q)\cos\big(k_r(z-z\pr+C)+\psi(k_r)\big)\rho_j(z\pr+m_j\lambda_1)dz\pr\ ,\nonumber\\
\label{vgt}\\
&&v^>(z)=\sum_{j=i+1}^{n_b}\xi_j\int_{\Omega_j}
\exp(-k_r(z-z\pr))/2Q)\cos\big(k_r(z-z\pr)+\psi(k_r))\big)\rho_j(z\pr+m_j\lambda_1)dz\pr\ ,\nonumber\\
\label{vlt}\\
&&v^d(z)=\xi_i\int_{z<z\pr}
\exp(-k_r(z-z\pr+C)/2Q)\cos\big(k_r(z-z\pr+C)+\psi(k_r))\big)\rho_i(z\pr+m_i\lambda_1)dz\pr\nonumber\\
&&\hskip 7mm +\xi_i\int_{z>z\pr}
\exp(-k_r(z-z\pr)/2Q)\cos\big(k_r(z-z\pr)+\psi(k_r))\big)\rho_i(z\pr+m_i\lambda_1)dz\pr\ .\nonumber\\
\label{vd}
\eea
The sums are regarded as empty when the lower limit exceeds the upper.

In each term of
(\ref{vgt}) and (\ref{vlt}) we  change the integration variable to $z_j=z\pr+m_j\lambda_1$, expand the cosine
by the double angle formula, and recognize the resulting integrals as real and imaginary parts of
a Fourier transform at $k_r$.
Then we find
\bea
&&v^<(z)=\nonumber\\&&\sum_{j=1}^{i-1}\xi_j\int_{\Omega_j} dz_j\rho_j(z_j)\exp(-k_r(z-z_j+m_j\lambda_1+C)/2Q)
\cos(k_r(z-z_j+m_j\lambda_1+C)+\psi(k_r))\nonumber\\
&&=\sum_{j=1}^{i-1}\xi_j\exp(-k_r(z+m_j\lambda_1+C)/2Q)\int_{\Omega_j} dz_j\exp(k_rz_j/2Q)\rho_j(z_j)\nonumber\\
&&\cdot\big[\cos\big(k_r(z+m_j\lambda_1+C)+\psi(k_r)\big)\cos(k_rz_j)+
\sin\big(k_r(z+m_j\lambda_1+C)+\psi(k_r)\big)\sin(k_rz_j)\big]\nonumber\\
&&=2\pi\sum_{j=1}^{i-1}\xi_j\exp(-k_r(z+m_j\lambda_1+C)/2Q)\nonumber\\ && \cdot\big[\cos\big(k_r(z+m_j\lambda_1+C)+\psi(k_r)\big)\rep\hat\rho_j(k_r)-
\sin\big(k_r(z+m_j\lambda_1+C)+\psi(k_r)\big)\imp\hat\rho_j(k_r)\big]\ ,\nonumber\\
\label{vltm}
\eea
where
\be
\hat\rho_j(k_r)=\frac{1}{2\pi}\int_{\Omega_j} e^{-k_rz(i-1/2Q)}\rho_j(z)dz\ .  \label{FT}
\ee
Carrying out a similar calculation for $v^>$, we may write the sum of the two terms as
\bea
&&v^<(z)+v^>(z)=2\pi\sum_{j=1}^{n_b}(1-\delta_{i,j})\xi_j\exp(-k_r(z+m_j\lambda_1+\theta_{i-1,j}C)/2Q)\nonumber\\
 &&\hskip 3cm \cdot\big[\cos\big(k_r(z+m_j\lambda_1+\theta_{i-1,j}C)+\psi(k_r)\big)\rep\hat\rho_j(k_r)\nonumber\\&&
 \hskip 3cm-
\sin\big(k_r(z+m_j\lambda_1+\theta_{i-1,j}C)+\psi(k_r)\big)\imp\hat\rho_j(k_r)\big]\ ,\quad
z\in\Omega_i\ , \label{vgtvlt}
\eea
where
\be
\theta_{i,j}=\bigg\{ \begin{array} {c} 1,~ i\ge j\\ 0,~ i < j \end{array}\ .   \label{thetadef}
\ee

The factor $\exp(k_rz/2Q)$ is very  close to 1 in the present application. By Table I we have $k_r/2Q=8\cdot 10^{-4}~{\rm m}^{-1}$, while $|z|$ is at most about $4\cdot 10^{-2}$~m in (\ref{FT}). Nevertheless, we shall not replace this factor by 1, since we expect to apply our formulas to cases with low $Q$ in future work.

The term $v^d$ of (\ref{vd}), representing the force on a bunch due to the field that it itself excited,
 is a bit more complicated. It is not an approximate  sinusoid, nor is it expressed in terms of the Fourier transform (\ref{FT}). It is, however, just one of $n_b$ terms, not distinguished in magnitude compared
to the others, because  the high $Q$ of the cavity implies long persistence of the fields excited. The various terms are distinguished principally in phase, through the phase shifts $m_j\lambda_1$ in their trigonometric arguments.
Barring unexpected cancellations, we can regard $v^d$ as a minor term, and try to determine it iteratively.

\section{Coupled Vlasov-Fokker-Planck and Ha\"issinski Equations \label{section:vfp}}
The kinetic equation for the phase space density of the $i$-th bunch is written in terms of the beam frame coordinate
$ z_i=z+m_i\lambda_1$ of that bunch . It is coupled to the equations for all other bunches through the induced voltage $V_r$, which depends
on the charge densities of all bunches. In view of the single-particle equations of motion (\ref{firstdenb})  (\ref{seconddenb}) , which
define the characteristic curves, the Vlasov-Fokker-Planck equation for the distribution function $f_i( z_i,\delta_i,t)$ of the $i$-th bunch takes the form
\be
\frac{\ptl f_i}{\ptl t} +\alpha c\delta_i\frac{\ptl f_i}{\ptl z_i}+\frac{1}{E_0T_0}\big(eV_1\sin(k_1 z_i+\phi_0)+eV_r( z_i-m_i\lambda_1)-U_0\big)
\frac{\ptl f_i}{\ptl\delta_i}=L^{FP}_if_i\ .    \label{vfp}
\ee
The Fokker-Planck term on the right hand side is
\be
L^{FP}_if_i=\frac{2}{\omega_st_d}\frac{\ptl}{\ptl\delta_i}\bigg[\delta_i f_i+\sigma_\delta^2\frac{\ptl f_i}{\ptl\delta_i}\bigg]\ ,
\label{fpdef}
\ee
where $\omega_s$ is the circular synchrotron frequency and $t_d$ is the longitudinal damping time.

We seek an equilbrium in which $\ptl f_i/\ptl t=0$ and $f_i$ has the factored Maxwell-Boltzmann form
\be
f_i(z_i,\delta_i)=\frac{1}{\sqrt{2\pi}\sigma_\delta}
\exp(-(\delta_i/\sigma_\delta)^2/2)~\rho_i( z_i)\ . \label{maxboltz}
\ee
Under this hypothesis the Fokker-Planck term vanishes and the spatial density $\rho_i$ must satisfy
\bea
&&\frac{d\rho_i}{d z_i}( z_i)=\frac{1}{\alpha c\sigma_\delta^2E_0T_0}\big(e\mathcal{V}( z_i )-U_0)~\rho_i( z_i)\ ,\nonumber\\
&& \mathcal{V}( z_i)=V_1\sin(k_1 z_i+\phi_0)+V_r( z_i-m_i\lambda_1) \label{haisde}
\eea
By separating variables and integrating we see that a solution must have the form
\be
\rho_i( z_i)=A_i^{-1} \exp\bigg[\frac{1}{\alpha c\sigma_\delta^2E_0T_0}
\int_0^{ z_i}\big(e\mathcal{V}(\zeta)-U_0\big)d\zeta\bigg]\ ,\quad -\Sigma\le z_i\le \Sigma\ ,\label{heqn}
\ee
where $A_i$ is a normalization constant, just the integral of the numerator over $[-\Sigma,\Sigma]$.

Noting that $cT_0=C$, we introduce the definitions
\be
\mu=\frac{1}{\alpha\sigma_\delta^2E_0C}\ ,\qquad U_i( z_i)=-\int_0^{ z_i}\big[e\mathcal{V}(\zeta)-U_0\big]d\zeta\ .  \label{muudef}
\ee
Recalling (\ref{haisde}) we then have (\ref{heqn}) expressed as
\be
\rho_i( z_i)=\frac{1}{A_i} \exp\big[-\mu U_i( z_i)\big]\ ,\quad A_i=\int_{-\Sigma}^\Sigma\exp\big[-\mu U_i(\zeta)\big]d\zeta\ ,
\quad -\Sigma\le z_i\le \Sigma\ ,\label{haissys}
\ee
where
\bea
&&U_i( z_i)=\frac{eV_1}{k_1}\big[\cos(k_1 z_i+\phi_0)-\cos\phi_0\big]+U_0 z_i\nonumber\\
&&+\frac{e^2N\omega_rR_s}{Q}\eta(k_r)\int_0^{ z_i}\big[v^<(\zeta-m_i\lambda_1)
+v^>(\zeta-m_i\lambda_1)+v^d(\zeta-m_i\lambda_1)\big]d\zeta\ .\nonumber\\ \label{Ui}
\eea
We refer to $U_i$ as the ``potential" for the $i$-th bunch, even though it has the dimension of an energy times a
length. The system of equations (\ref{haissys}) for $i=1,\cdots,n_b$ will be called the {\it coupled Ha\" issinski equations}.
\section{Mean energy transfer in the equilibrium state \label{section:mean_E_trans}}
According to (\ref{firstdenb}) the power transferred to a single particle with coordinate $z_i$ in the $i$-th bunch is
\be
P_i(z_i)=dE_i/dt=E_0d\delta_i/dt=\frac{1}{T_0}(eV_1\sin(k_1z_i+\phi_0)+eV_r(z_i-m_i\lambda_1)-U_0)\ .\label{single_power}
\ee
The mean value of the power over the equilibrium distribution is obtained from (\ref{haisde}) as
\be
\int \rho_i(z_i)P_i(z_i)dz_i=\alpha c\sigma_\delta^2T_0 \int \frac{d\rho_i}{dz_i}dz_i=0\ ,
\ee
thus for every $i$,
\be
U_0=U_1+U_r=eV_1\int \sin(k_1z+\phi_0)\rho_i(z)dz\ +e\int V_r(z-m_i\lambda_1)\rho_i(z)dz\ .\label{balance}
\ee
The first term on the right hand side is the mean energy supplied by the external r.f., while the second term, which is
negative, represents the mean energy lost to the harmonic cavity per turn. We automatically have energy balance, on the average, in the
equilibrium state.

\section{Integral of the induced voltage \label{section:integral}}

 To express the integral in (\ref{Ui}) we define $\mathcal{S}$ and $\mathcal{C}$ as follows:
\bea
&&\mathcal{S}(k_rz,\phi)=k_r\int_0^z\exp(-k_r\zeta/2Q)\cos(k_r\zeta+\phi)d\zeta\nonumber\\
&&=\bigg[\frac{\exp(-k_rz/2Q)}{1+1/4Q^2}
\big[\sin(k_rz+\phi)-\frac{1}{2Q}\cos(k_rz+\phi)\big]\bigg]_0^z\nonumber\\
&&\approx\exp(-k_rz/2Q)\sin(k_rz+\phi)-\sin\phi\ ,\label{expcos}\\
&&\mathcal{C}(k_rz,\phi)=-k_r\int_0^z\exp(-k_r\zeta/2Q)\sin(k_r\zeta+\phi)d\zeta \nonumber\\&&=\bigg[\frac{\exp(-k_rz/2Q)}{1+1/4Q^2}
\big[\cos(k_rz+\phi)+\frac{1}{2Q}\sin(k_rz+\phi)\big]\bigg]_0^z\nonumber\\
&&\approx \exp(-k_rz/2Q)\cos(k_rz+\phi)-\cos\phi\ .\label{expsin}
\eea
The large-$Q$ approximation stated here on the right is not used in our code, since we wish to be
set up for later applications with small $Q$

Applying this in (\ref{vgtvlt}) and  we find
\bea
&&\int_0^{ z_i}\big[v^<(\zeta-m_i\lambda_1)+v^>(\zeta-m_i\lambda_1)\big]d\zeta= \nonumber\\&&\frac{2\pi}{k_r}\sum_{j=1}^{n_b}(1-\delta_{i,j})\xi_j
\exp(-\phi_{i,j}/2Q)\cdot\big[\mathcal{S}(k_r z_i,\phi_{i,j}+\psi)\rep\hat\rho_j
+\mathcal{C}(k_r z_i,\phi_{i,j}+\psi)\imp\hat\rho_j\big] \hskip 1cm\label{intvltvgt}\\
&&\hskip 3cm \phi_{i,j}=k_r((m_j-m_i)\lambda_1+\theta_{i-1,j}C)\ .\label{phij}
\eea

It remains to calculate
\be
\int_0^{ z_i}v^d(\zeta-m_i\lambda_1)d\zeta\ ,   \label{intvd}
\ee
with $v^d$ from (\ref{vd}).  After changing  the integration variable in  (\ref{vd}) to $u=z\pr+m_i\lambda_1$ we have
\bea
&&v^d(\zeta-m_i\lambda_1)=\xi_i\int_{-\Sigma}^{\zeta}\exp(-k_r(\zeta-u)/2Q)\cos(k_r(\zeta-u)+\psi)\rho_i(u)du\nonumber\\
&&+\xi_i\int_{\zeta}^\Sigma\exp(-k_r(\zeta-u+C)/2Q)\cos(k_r(\zeta-u+C)+\psi)\rho_i(u)du \label{vdm}
\eea
We can avoid the double integral in (\ref{intvd}) through an integration by parts. After applying the double angle formula to the
cosine, one of the terms comprising (\ref{intvd}) takes the form
\be
\xi_i\int_0^{z_i}\exp(-k_r\zeta/2Q)\cos(k_r\zeta+\psi)\int_{-\Sigma}^\zeta\exp(k_ru/2Q)\cos(k_ru)\rho_i(u)du\ .
\ee
Now in a partial integration the factor $\exp(-k_r\zeta/2Q)\cos(k_r\zeta+\psi)$ is integrated by applying (\ref{expcos}) , while the $u$-integral
is differentiated. Proceeding similarly with the other terms, we eliminate all double integrals.

\section{Solution of coupled Ha\" issinski equations by Newton's method \label{section:newton}}
Let us multiply (\ref{haissys}) by $\exp(k_rz_i/2Q)$ and then take the Fourier transform, as in (\ref{FT}). This yields
\be
\hat\rho_i(k_r)-\frac{1}{2\pi A_i }\int_{-\Sigma}^\Sigma \exp\big[-k_rz(i-1/2Q)-\mu U_i(z)~\big]dz=0\ ,
\quad i=1,\cdots,n_b \label{maineq}
\ee
where $A_i$ is defined in (\ref{haissys}) and
\bea
&&U_i(z)=\frac{eV_1}{k_1}\big[\cos(k_1z+\phi_0)-\cos\phi_0\big]+U_0z+\frac{e^2N\omega_rR_s\eta}{Q}\bigg[
\int_0^zv^d(\zeta-m_i\lambda_1)d\zeta\nonumber\\&&
\frac{2\pi}{k_r}\sum_{j=1}^{n_b}(1-\delta_{i,j})\xi_j
\exp(-\phi_{i,j}/2Q)\cdot\big[\mathcal{S}(k_r z_i,\phi_{i,j}+\psi)\rep\hat\rho_j
+\mathcal{C}(k_r z_i,\phi_{i,j}+\psi)\imp\hat\rho_j\big] \hskip 1cm\label{usubi}
\eea

If the diagonal term in $v^d$ were known, the real and imaginary parts of (\ref{maineq}) constitute
$2n_b$ equations in the $2n_b$ unknowns $\rep\hat\rho_j,  \imp\hat\rho_j$. Defining a notation for the
diagonal term,
\be
u^d_i(z)=\int_0^zv^d(\zeta-m_i\lambda_1)d\zeta\ ,   \label{uddef}
\ee
we write  (\ref{maineq}) more briefly as
\be
F(\hat\rho,u^d)=0 \ . \label{fzero}
\ee
where $F$ and $\hat\rho$ are complex column vectors with $n_b$ components, in which
\be
F_i(\hat\rho,u^d)=A_i\hat\rho_i-\frac{1}{2\pi}\int_{-\Sigma}^\Sigma \exp\big[-k_rz(i-1/2Q)-\mu U_i(z,\hat\rho,u^d)~\big]dz
\ee

 For given $u^d$ we try to
solve (\ref{fzero}) by the matrix form of Newton's method, namely
\be
F(\hat\rho^{(n)},u^d)+\frac{\ptl F(\hat\rho^{(n)}, u^d)}{\ptl\rep\hat\rho}\rep(\hat\rho^{(n+1)}-\hat\rho^{(n)} )
+\frac{\ptl F(\hat\rho^{(n)}, u^d)}{\ptl\imp\hat\rho}\imp(\hat\rho^{(n+1)}-\hat\rho^{(n)} )=0\ .  \label{newton}
\ee
Here $(\ptl F/\ptl\rep\hat\rho,~\ptl F/\ptl\imp\hat\rho)$ are complex matrices  with elements $(\ptl F_i/\ptl\rep\hat\rho_j,~\ptl F_i/\ptl\imp\hat\rho_j)$.
That is, we linearize $F$ about iterate $\hat\rho^{(n)}$ to define the update $\hat\rho^{(n+1)}$ by solving $2n_b$ real
linear equations for the increment $\hat\rho^{(n+1)}-\hat\rho^{(n)}$.

 Lacking any better choice, we begin the process with $\hat\rho^{(0)}$ obtained from Gaussians, all with the nominal bunch length:
\be
\hat\rho^{(0)}_i(k_r)=\frac{1}{(2\pi)^{3/2}\sigma_z}\int_{-\Sigma}^\Sigma \exp\big(k_rz(-i+1/2Q)-(z/\sigma_z)^2/2\big)dz\ ,\quad i=1,\cdots,n_b   \label{gaussstart}
\ee

Note that we could not use the direction-independent complex derivative $\ptl/\ptl\hat\rho_i$, since $U_i$ is not an analytic function of $\hat\rho_i$, being always real.

To account for the diagonal term we adopt the simple device of computing $u^d$ in (\ref{newton}) from the previous Newton iterate. That procedure yields a convergent scheme, and shows that the contribution of the diagonal term
is negligible, at least in the present case of a high-$Q$ cavity. If our scheme is later applied to a low-$Q$ case, a more sophisticated
method might be needed to determine $v^d$.

\section{Expression of the Jacobian matrix \label{section:jacobian}}
Since the exponent $U_i$ is linear in the unknowns $\hat\rho_j$, it is not difficult to write down the Jacobian, the
matrix of the partial derivatives that appear in (\ref{newton}). One must not forget the derivatives of $A_i$, which
are essential to ensure that the final $\rho_i(z_i)$ are automatically normalized to have unit integral. The complete
$2n_b\times 2n_b$ Jacobian in block matrix form is
\be
\left[\begin{array} {cc} \ptl\rep F/\ptl\rep\hat\rho &  \ptl\rep F/\ptl\imp\hat\rho\\
\ptl\imp F/\ptl\rep\hat\rho & \ptl\imp F/\ptl\imp\hat\rho\end{array}\right]
\ ,  \label{blockJ}
\ee
where
\bea
&&\frac{\ptl\rep F_i}{\ptl\rep\hat\rho_j}=A_i\delta_{ij}+\mu\int\bigg[\rep\hat\rho_i+\frac{1}{2\pi}
\cos (k_rz) e^{k_rz/2Q}\bigg]e^{-\mu U_i(z)}a_{ij}(z)s_{ij}(z)dz\ ,\label{rere}\\
&&\frac{\ptl\rep F_i}{\ptl\imp\hat\rho_j}=\mu\int\bigg[\rep\hat\rho_i+\frac{1}{2\pi}
\cos (k_rz) e^{k_rz/2Q}\bigg]e^{-\mu U_i(z)}a_{ij}(z)c_{ij}(z)dz\ ,\label{reim}\\
&&\frac{\ptl\imp F_i}{\ptl\rep\hat\rho_j}=\mu\int\bigg[\imp\hat\rho_i-\frac{1}{2\pi}
\sin (k_rz) e^{k_rz/2Q}\bigg]e^{-\mu U_i(z)}a_{ij}(z)s_{ij}(z)dz\ ,\label{imre}\\
&&\frac{\ptl\imp F_i}{\ptl\imp\hat\rho_j}=A_i\delta_{ij}+\mu\int\bigg[\imp\hat\rho_i-\frac{1}{2\pi}
\sin (k_rz) e^{k_rz/2Q}\bigg]e^{-\mu U_i(z)}a_{ij}(z)c_{ij}(z)dz\ ,\label{imim}
\eea
with
\bea
&&a_{ij}(z)=\frac{2\pi e^2N\omega_rR_s\eta}{k_rQ}\big(1-\delta_{ij}\big)\xi_j
\exp\big[-k_r\big((m_j-m_i)\lambda_1+\theta_{i-1,j}C\big)/2Q\big]\ ,\label{aij}\\
&&s_{ij}(z)=\mathcal{S}(k_r( z+(m_j-m_i)\lambda_1+\theta_{i-1,j}C),\psi)\ ,\label{sij}\\
&&c_{ij}(z)=\mathcal{C}(k_r( z+(m_j-m_i)\lambda_1+\theta_{i-1,j}C),\psi)\ .\label{cij}
\eea
\section{Continuation in current \label{section:current}}

In contrast to experience with the Ha\" issinski equation for a single bunch, we shall find that the
Newton iteration (\ref{newton}) beginning with (\ref{gaussstart}) does not converge at the desired design current. This must be because the solution at full current deviates
extremely from the unperturbed Gaussian, whereas the deviation is relatively small in the single bunch case.

At small current the Jacobian is nearly diagonal and positive definite, since the off-diagonal terms have a
factor $eN$. This augurs well for the success of the Newton iteration at sufficiently small current. It then
seems reasonable to get a solution  with small current, then take that solution as the starting point for a
Newton iteration at a somewhat higher current. If the second iteration converges we can perhaps repeat the process several times to reach the required large current.

The calculation could be made more efficient by extrapolating linearly in current after each successful iteration.
This should allow a larger increment in current.
Let us define a convenient current parameter such as $I=I_{avg}$, the average bunch current.
 Expanding the notation to include
$I$-dependence, and suppressing reference to $u^d$, we write (\ref{fzero}) as
\be
F(\hat\rho(I),I)=0\ , \label{finci}
\ee
and differentiate with respect to $I$ to obtain
\be
\frac{\ptl F}{\ptl\rep\hat\rho}\frac{d\rep\hat\rho}{dI} +\frac{\ptl F}{\ptl\imp\hat\rho}\frac{d\imp\hat\rho}{dI}
+\frac{\ptl F}{\ptl I}= 0\ .  \label{drhodi}
\ee
The solution of this linear system for $d\hat\rho/dI$ affords the linear extrapolation
\be
\hat\rho(I+\Delta I)=\hat\rho(I)+\frac{d\hat\rho(I)}{dI}\Delta I\ .   \label{Iextrap}
\ee
The extrapolation is not a costly step, since the Jacobian matrix in (\ref{drhodi}) is already known from the previous Newton iteration.

It is helpful to redefine the unknown as $\tilde\rho=I\hat\rho$, so that the current appears linearly in the transformed version of (\ref{finci}), namely
\be
\tilde F_i(\tilde\rho,I)=A_i\tilde\rho_i-\frac{I}{2\pi}\int_{-\Sigma}^\Sigma \exp\big[-k_rz(i-1/2Q)-\mu \tilde U_i(z,\tilde\rho)~\big]dz=0\ .
\ee
 Now even the right hand side of the equation to be solved, $-\ptl\tilde F/\ptl I$, is an integral already
 computed during the previous Newton iteration.

Our continuation procedure is an example of a general method for solving nonlinear problems, called {\it path following} or {\it executing a homotopy} \cite{gene}. One follows a known or easily computed solution as a function of a
parameter, which could be multidimensional. The continuation could stall or display a bifurcation if a singularity
of the Jacobian were encountered.
\section{Numerical results for parameters of ALS-U \label{section:numerical}}
\subsection{Complete train without gaps \label{subsection:complete}}
We first present results for a complete train without gaps, for which $n_b=h=328$, at the nominal current of $500$~mA.
The calculation starts at a current of $150$~mA and proceeds to the desired current in three equal increments, by means of the algorithm of the previous section. The convergence criterion for a Newton iteration is in terms of a sum
of normalized residuals of the equation (\ref{fzero}) that is to be satisfied:
\be
\epsilon=\sum_{i=1}^{n_b}\frac{|\rep F_i|+|\imp F_i|}{A_i(|\rep\hat\rho_i|+|\imp\hat\rho_i|)}
\ee
With convergence defined by  $\epsilon < 10^{-12}$, the Newton iterations converge in at most 7 steps, and the CPU time for the whole calculation is 25 seconds on a laptop PC. The program is a  serial code in Fortran, using the Intel Math Kernel Library for the linear algebra. As is typical of Newton's method, the convergence is very rapid at the last steps, with $\epsilon$ bounded by $\epsilon^2$ of the previous step. The number of mesh points for the integrals on $z$ was 201, and the mesh covers $[-6\sigma_{z0},\ 6\sigma_{z0}]$. The solution with twice as many points is the same to graphical accuracy, and takes twice the CPU time.

\begin{figure}[htb]
   \centering
   \includegraphics[width=\linewidth]{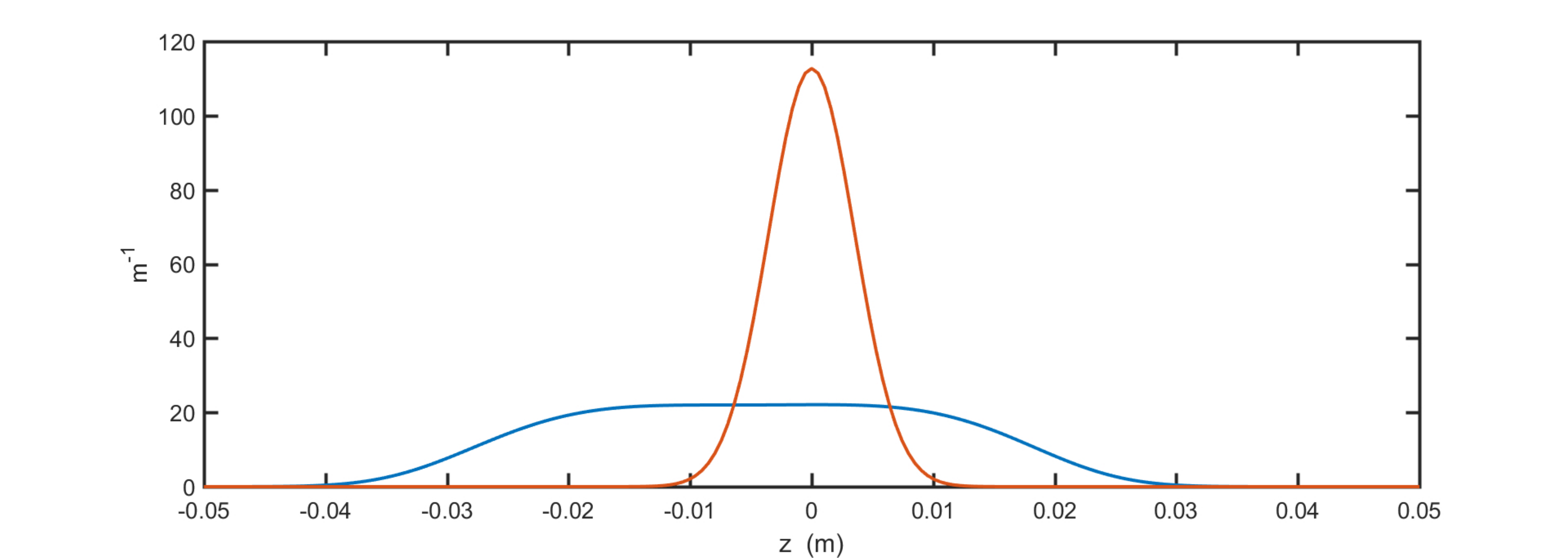}
  \caption{Charge density for complete fill at 500 mA (blue), with r.m.s. width $\sigma=4.08\sigma_{z0}$ and centroid $<z>=-4.74~$mm. Gaussian solution at low current (red) with $\sigma_{z0}=3.54~$mm.}
   \label{fig: fig3}
\end{figure}

Figure \ref{fig: fig3} shows the resulting charge density of $328$ identical bunches, for the parameters of Table I, together with the Gaussian low current solution having the natural bunch width $\sigma_{z0}=3.54$~mm. At
full current the harmonic cavity increases the r.m.s. width to $4.08\sigma_{z0}$, and causes a centroid displacement
of $-4.74$~mm.

\begin{figure}[htb]
   \centering
   \includegraphics[width=\linewidth]{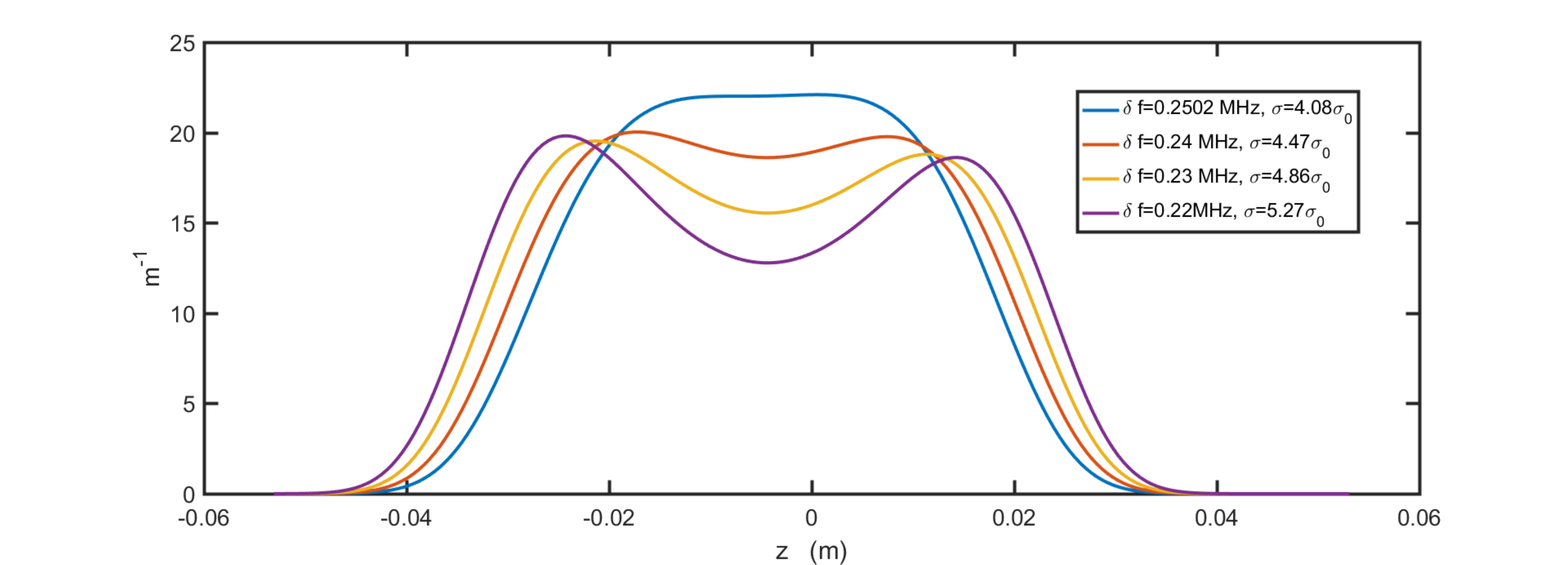}
  \caption{Charge density for complete fill at 500 mA, for decreasing values of the detuning $\delta f=$\\$f_r-3f_1=(.2502,~ .24, ~.23, ~.22) $~MHz, with corresponding bunch lengths $\sigma=(4.08, ~4.47,~ 4.86,~ 5.27)~\sigma_0$.} 
   \label{fig: fig4}
\end{figure}

The shunt impedance and detuning in Table I were chosen to maximize the bunch lengthening at the nominal current, while keeping a flat top in the density. At higher impedance or lower detuning one can achieve a larger bunch
lengthening, but at the expense of getting a density with two maxima. The bunch in this situation is sometimes described
as being ``over-stretched". Figure \ref{fig: fig4} shows the effect of
decreasing the detuning, and lists the corresponding r.m.s bunch lengths.


It is important to note that all bunch forms turn out to be the same, merely by putting $n_b=h$, even though the
equations contain no explicit constraint that they be the same. This is gratifying and as it should be by physical intuition, but the mathematical mechanism for it to happen is somewhat obscure.

\subsection{Train with a single gap \label{subsection:single}}
\begin{figure}[htb]
   \centering
   \includegraphics[width=.8\linewidth, height=.35\linewidth]{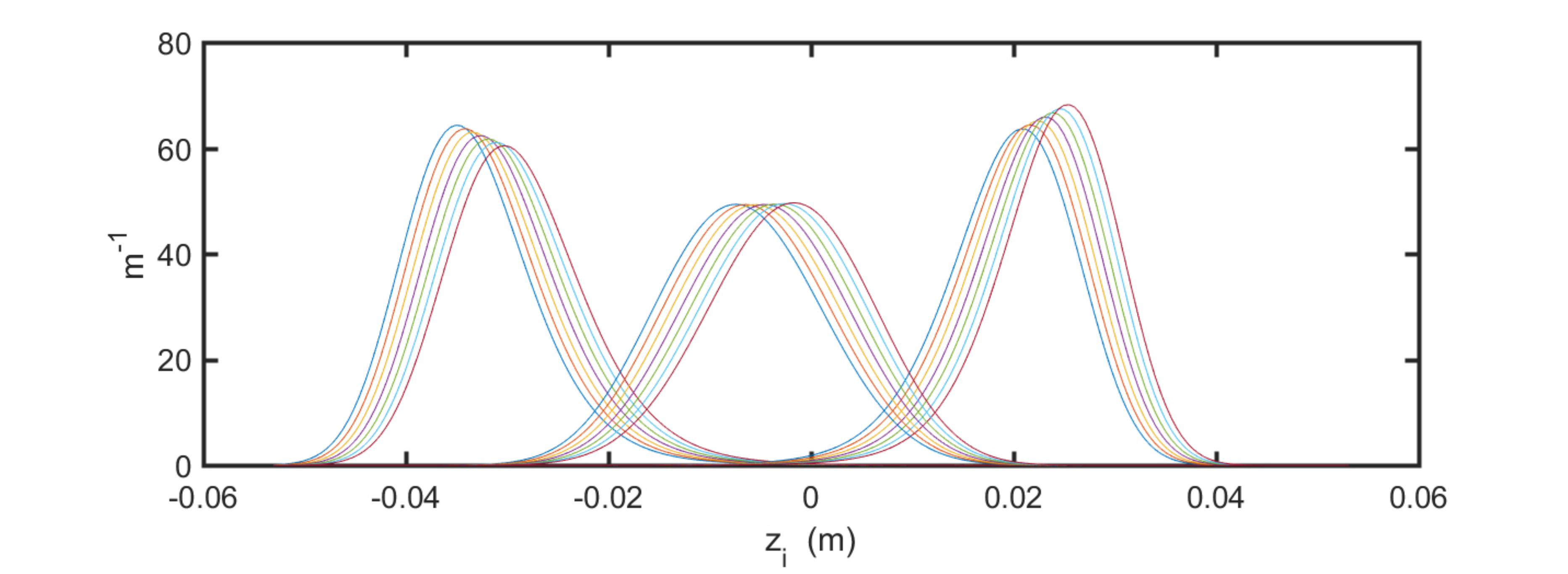}
  \caption{Results for a single gap of 44 empty buckets. On the right, 7 bunches within the first 24; in the middle,
  7 bunches within the middle 24; on the left, 7 bunches within the last 24.}
   \label{fig: fig5}
\end{figure}
\begin{figure}[htb]
   \centering
   \includegraphics[width=\linewidth]{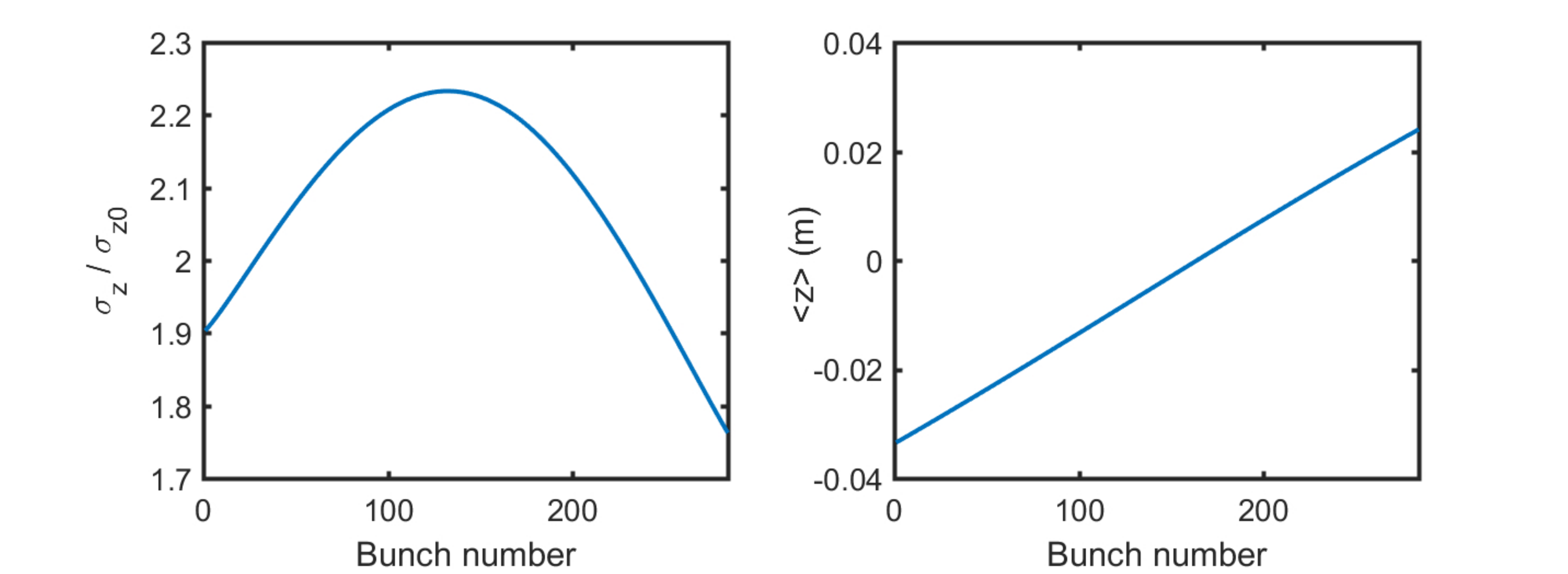}
  \caption{Results for a single gap of 44 empty buckets. In the left plot, the ratio of r.m.s. bunch length to the natural  bunch length vs. bunch number, head of train on the right. In the right plot, the corresponding graph of the centroid position.}
   \label{fig: fig6}
\end{figure}
Next we consider a train with a single gap of 44 empty buckets, thus 284 filled buckets in a row. The average current,
i.e., the total charge divided by the revolution time, is taken to be the same as before, corresponding
to the individual bunch charge being larger by a factor 328/284. With an initial average current of 150~mA, increased to 500~mA in three steps, the convergence is even better than in the previous example.

In Figure \ref{fig: fig5} we show representative bunch forms near the front, middle, and end of the train. Each bunch
is given as a function of its beam frame coordinate $z_j=z+m_j\lambda_1$.  There is
much less bunch lengthening than in the complete fill, and a large centroid shift varying linearly along the train. In Figure
\ref{fig: fig6} (left) we show the variation of bunch length (divided by the natural bunch length) along the train, while Figure \ref{fig: fig6} (right) shows the variation of the centroid position. The maximum centroid shift is 7 times
larger than in the complete fill.

\subsection {Train with distributed gaps \label{subsection:distributed}}
The sharp reduction in bunch lengthening induced by a single gap leads to the idea of distributing the empty buckets around the ring as much as possible \cite{pan}. This has a chance of resembling more closely the complete fill. For ALS-U the minimum acceptable gap consists of 4 empty buckets, since a gap of 10ns is required to accommodate the rise and fall times of the fast kicker that does on-axis injection from the accumulator ring.  With such gaps we need 9 trains of 26 bunches and two of 25 to account for 328 buckets total: $9\times 26+2\times 25+11\times 4=328$.
\begin{figure}[htb]
   \centering
   \includegraphics[width=.8\linewidth, height=.35\linewidth]{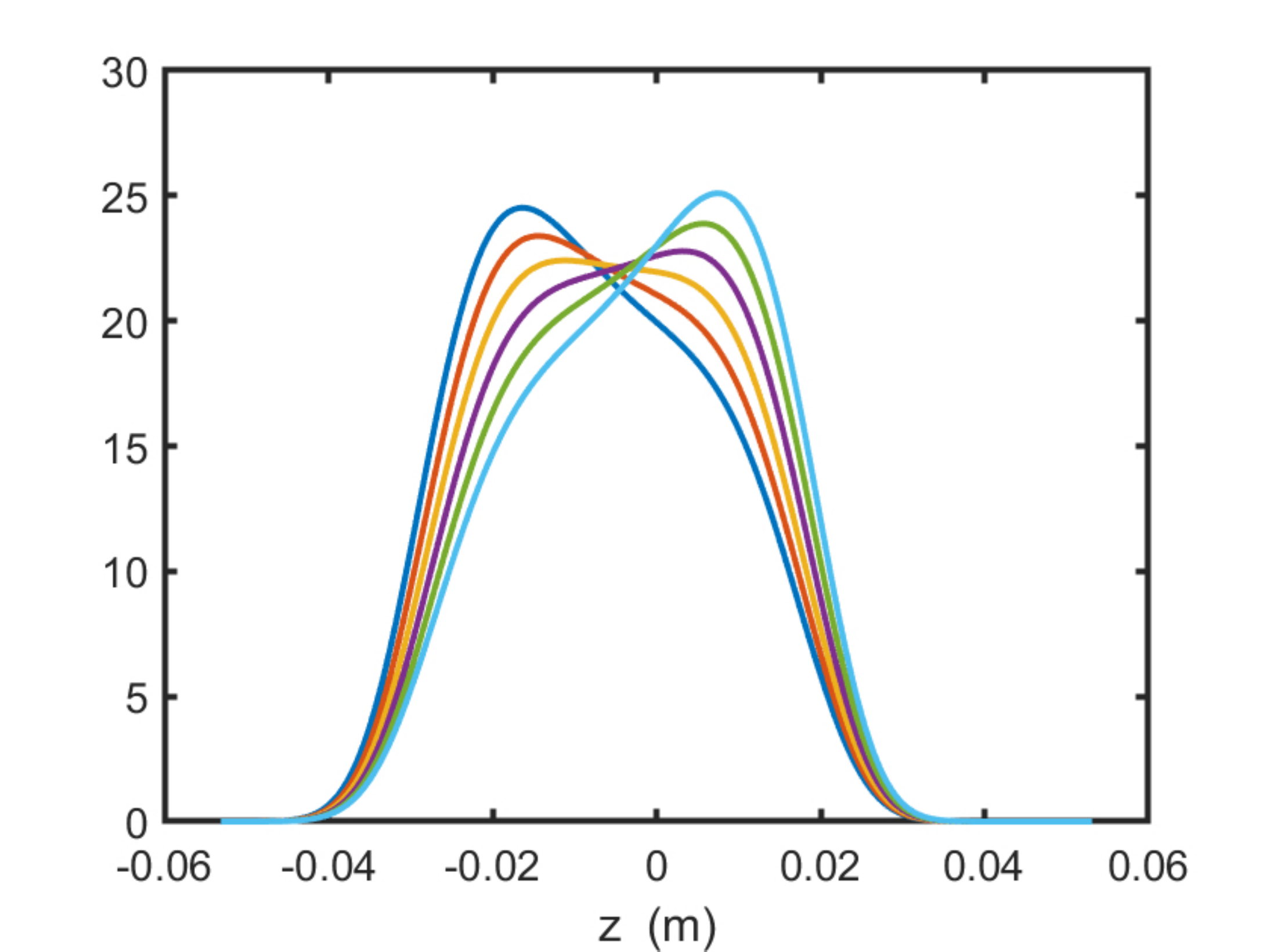}
  \caption{Results for distributed gaps, each consisting of 4 empty buckets. Plots of 6
  bunches out of a train of 26. }
   \label{fig: fig7}
\end{figure}
\begin{figure}[htb]
   \centering
   \begin{minipage} [b]{.49\linewidth}
   \includegraphics[width=\linewidth]{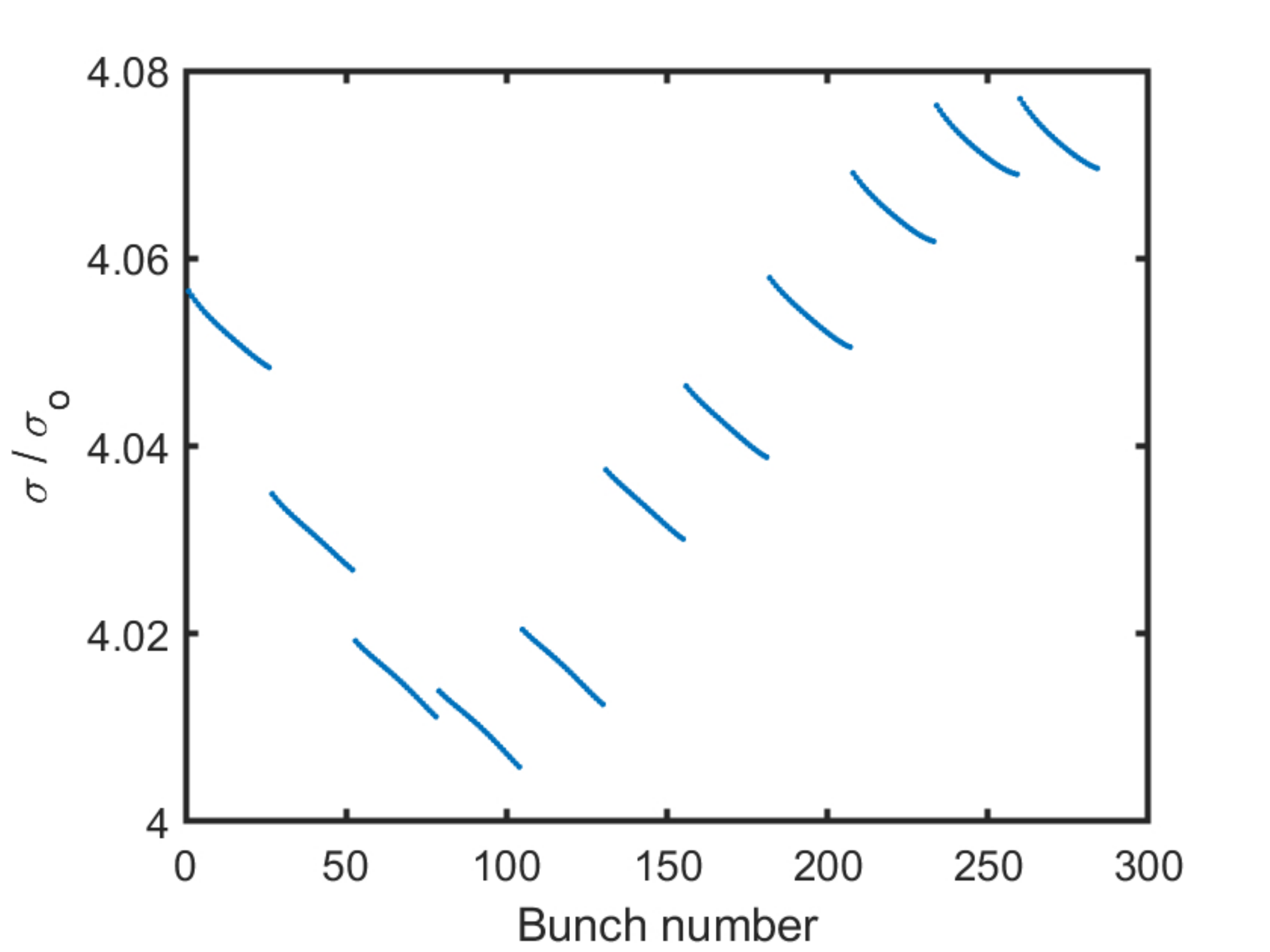}
  \caption{Ratio of bunch length to natural \\ bunch length vs. bunch number, with\\ distributed gaps.}
   \label{fig: fig8}
   \end{minipage}
   \begin{minipage} [b]{.49\linewidth}
   \includegraphics[width=\linewidth]{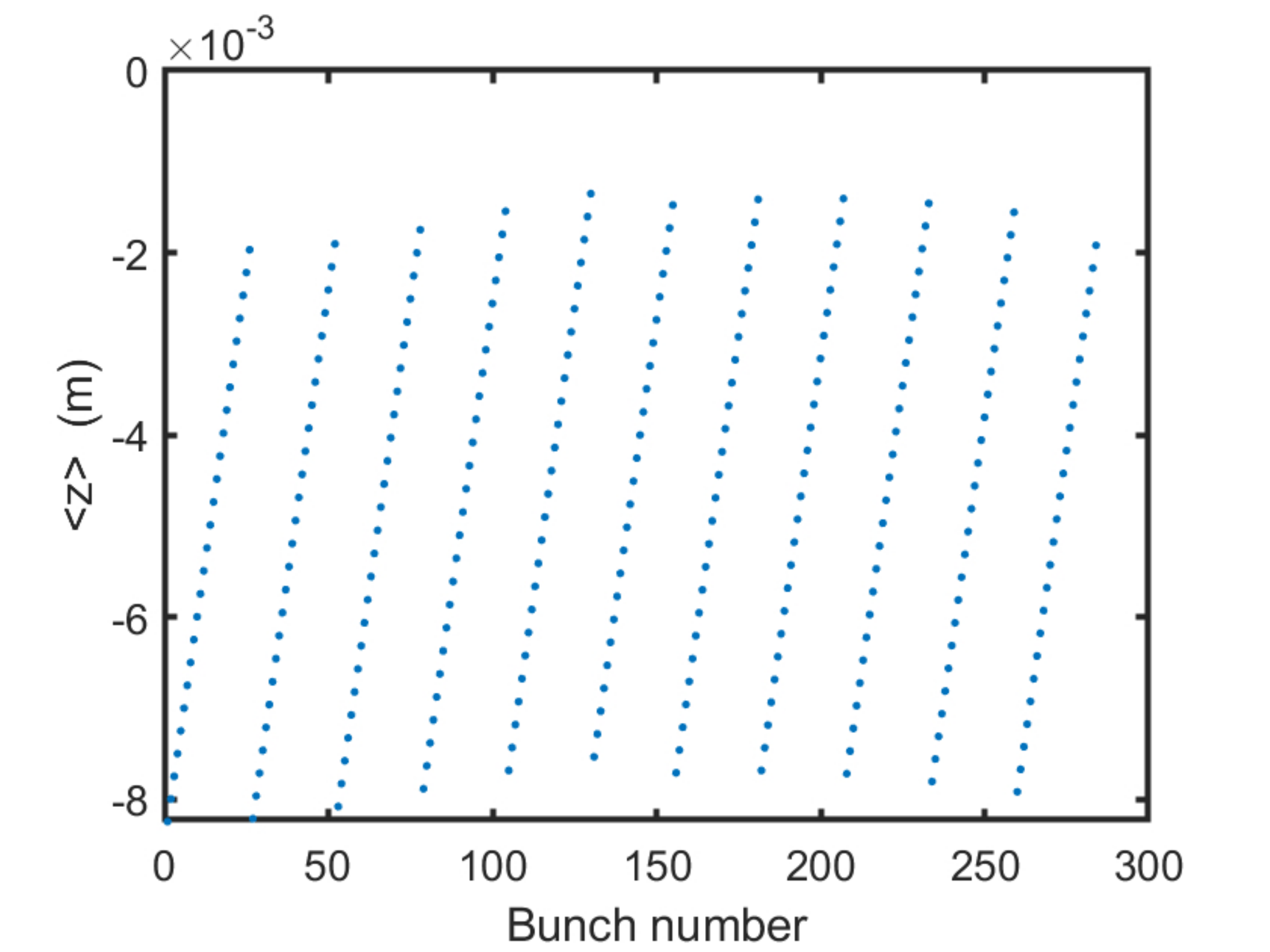}
   \caption{Centroid vs. bunch number, with distributed gaps.\\}
   \label{fig: fig9}
   \end{minipage}
\end{figure}

We consider the Case C2 of Ref.~\cite{pan}, in which the two trains of 25 are as far apart as possible. This was found to be slightly more helpful than putting those two side-by-side. Figure \ref{fig: fig7} shows the result for 6 bunches out of a train of 26, including the initial and final bunches. Figure \ref{fig: fig8} shows the ratio of the r.m.s. bunch length to the natural bunch length vs. bunch number, while Figure \ref{fig: fig9} displays the centroid vs. bunch number. Fortunately, the average bunch lengthening now has a value near the case of the complete fill.
Furthermore, the big centroid displacement of the single gap case is gone. There is a small and linear centroid displacement along each sub-train, but its magnitude is similar to that of the complete fill.

Although maximal distribution of the gaps is a step in the right direction, it leaves us with a strong variation of bunch form along the train
and some highly skewed charge distributions. We should then look for further means to imitate the complete fill as much as possible.

  \subsection{Guard bunches to compensate the damage from gaps. \label{subsection:guards}}
  \begin{figure}[htb]
   \centering
   \includegraphics[width=.8\linewidth, height=.35\linewidth]{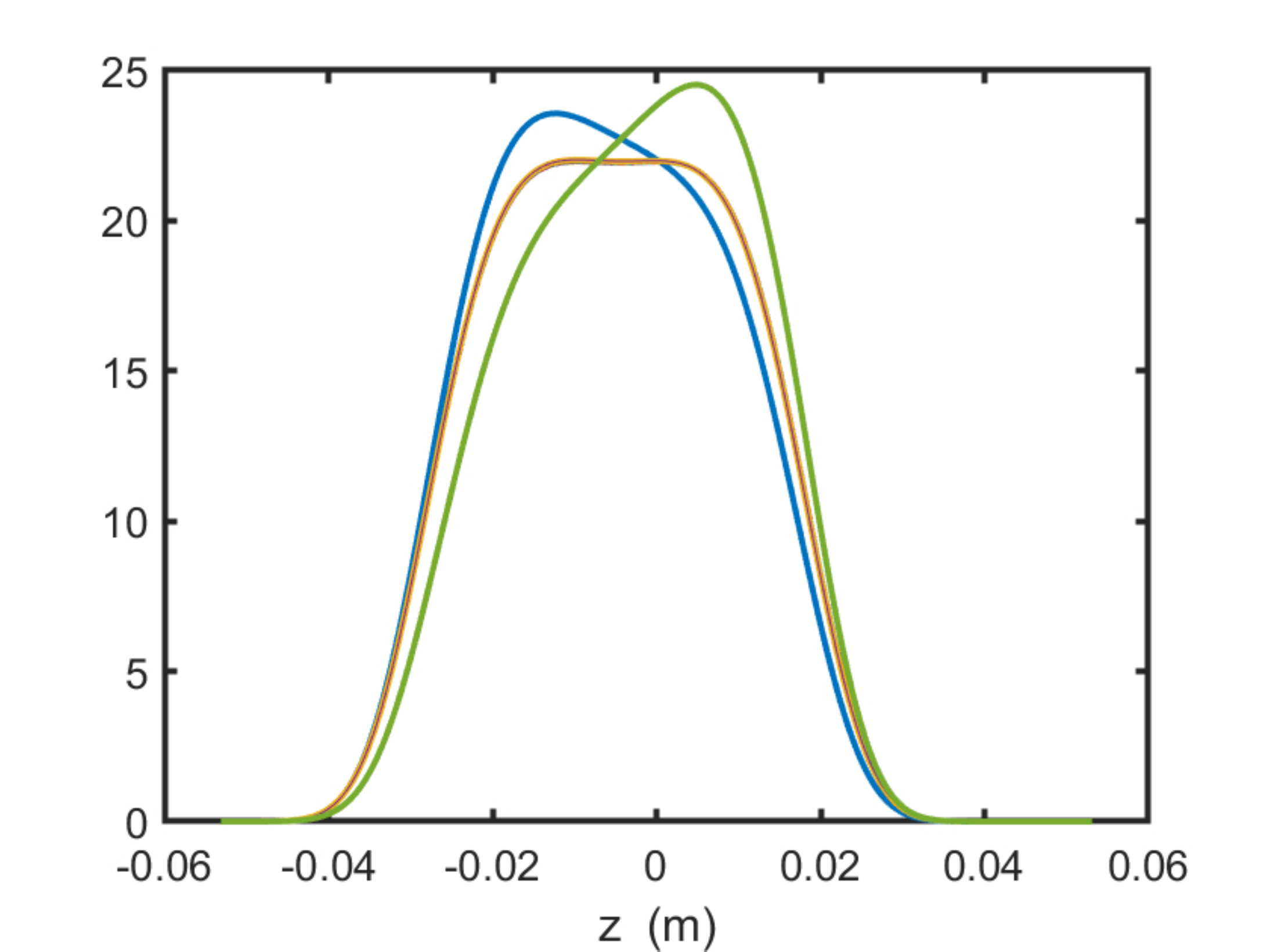}
  \caption{A train of 26 with a guard bunch at its beginning (on the right) and at its end (on the left). The guard
  bunches have 3 times the charge of the inner bunches, to compensate for the charge missing in 4 gaps. The 24 inner bunches (in red)
  are nearly identical.}
   \label{fig: fig10}
\end{figure}
\begin{figure}
  \begin{minipage}[b]{0.49\linewidth}
  \includegraphics[width=\linewidth]{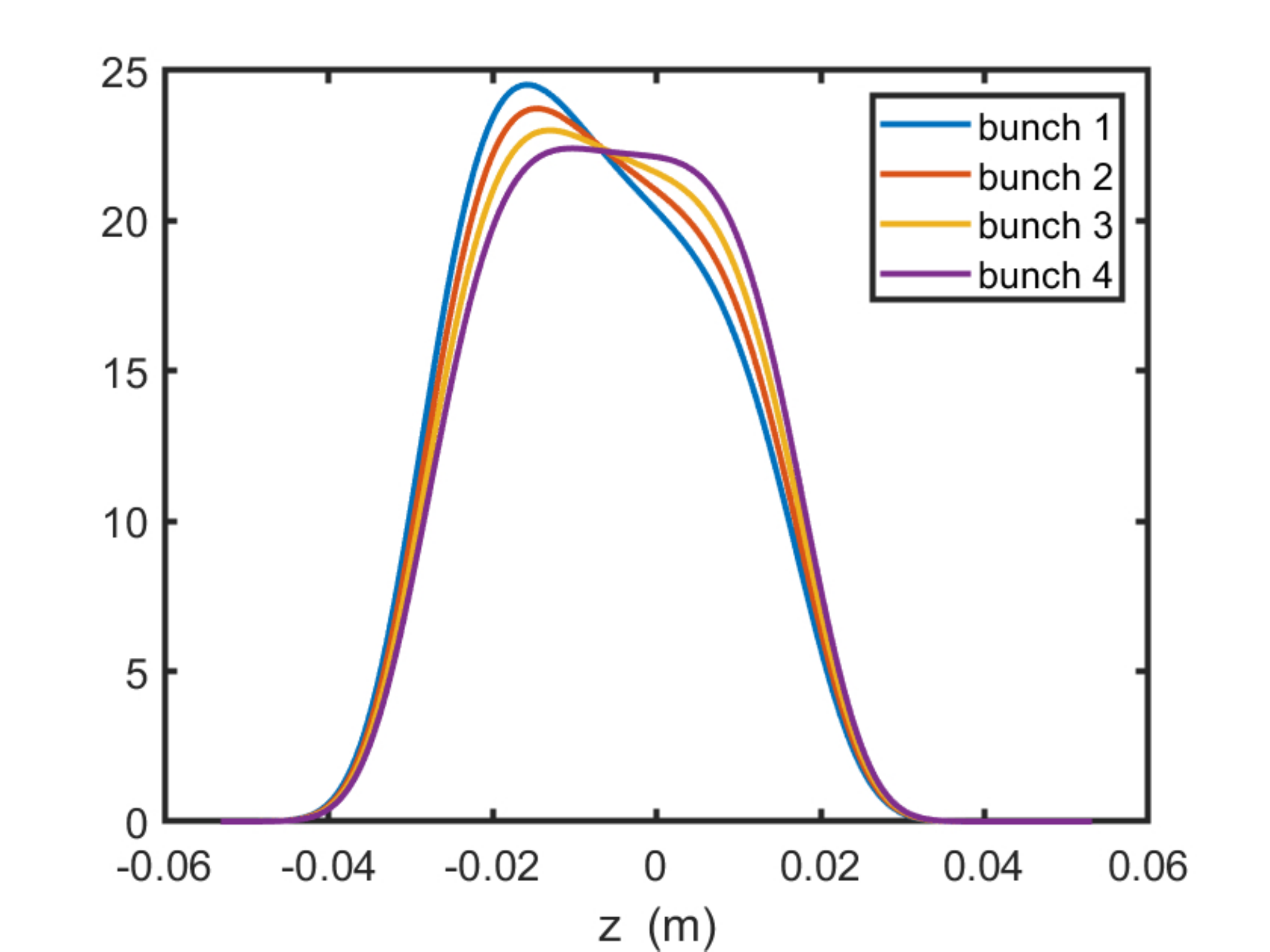}
  \caption{The 4 guard bunches at the end of a\\ train of 26, each with $\xi=1.5$. These lean\\
  forward.}
  \label{fig: fig11}
  \end{minipage}
   \begin{minipage}[b]{0.49\linewidth}
  \includegraphics[width=\linewidth]{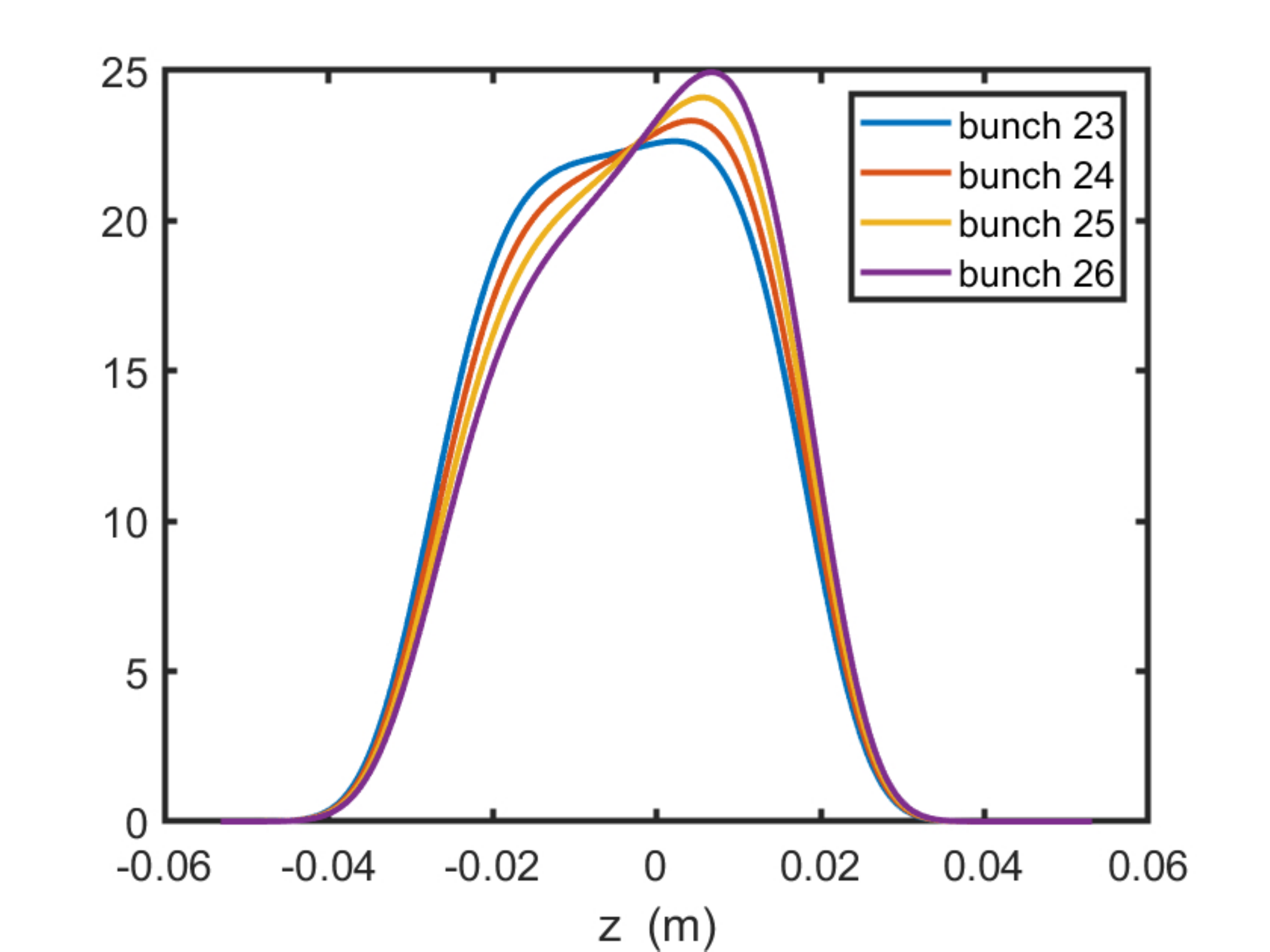}
  \caption{The 4 guard bunches at the beginning of a train of 26, each with $\xi=1.5$. These
  lean backward.}
  \label{fig: fig12}
  \end{minipage}
   \begin{minipage}[b]{0.49\linewidth}
  \includegraphics[width=\linewidth]{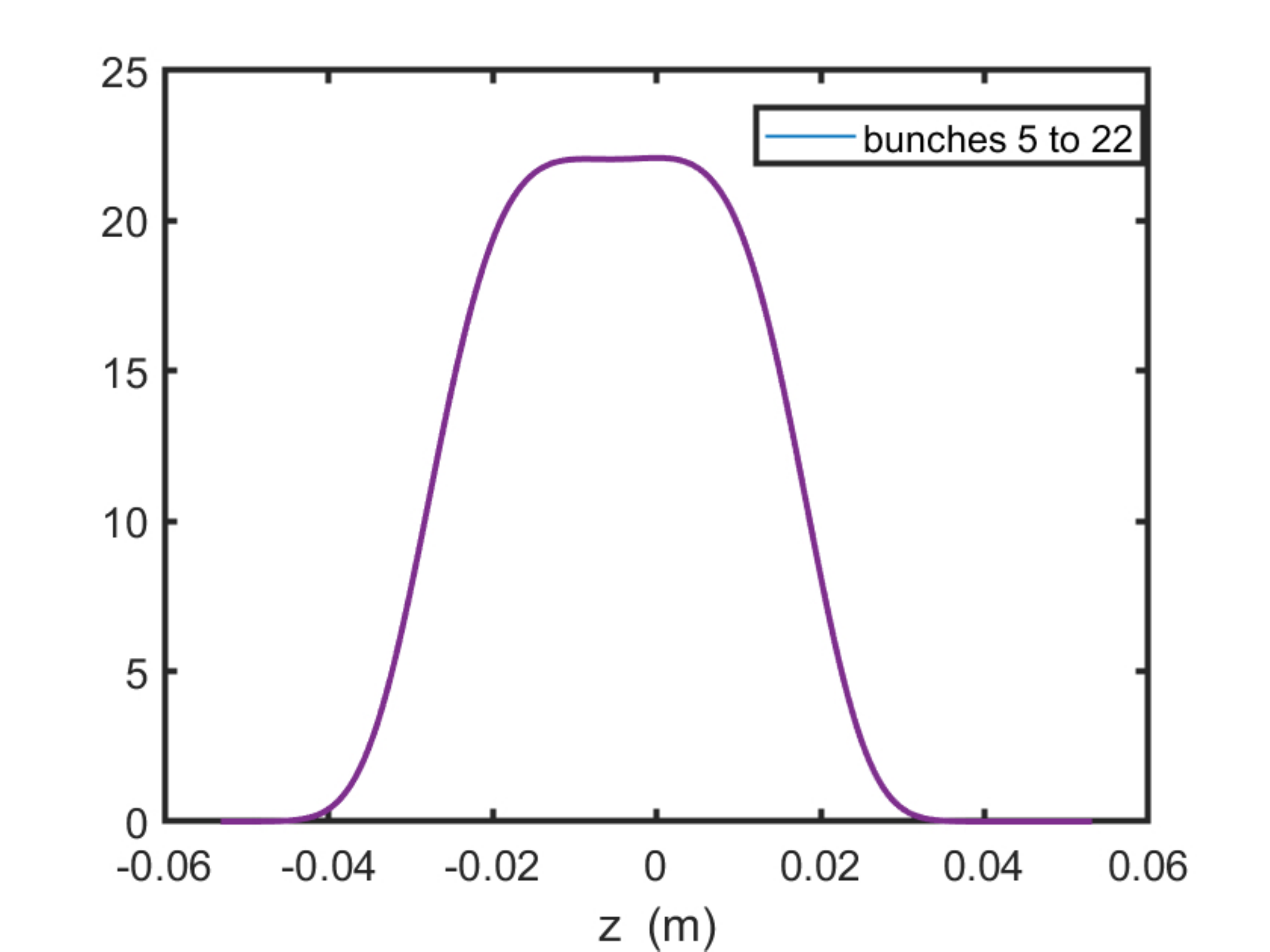}
  \caption{The 18 normal bunches in the middle of the train, each with $\xi=1$.}
  \label{fig: fig13}
  \end{minipage}
  \end{figure}
    \begin{figure}[htb]
   \centering
   \begin{minipage} [b]{.49\linewidth}
   \includegraphics[width=\linewidth]{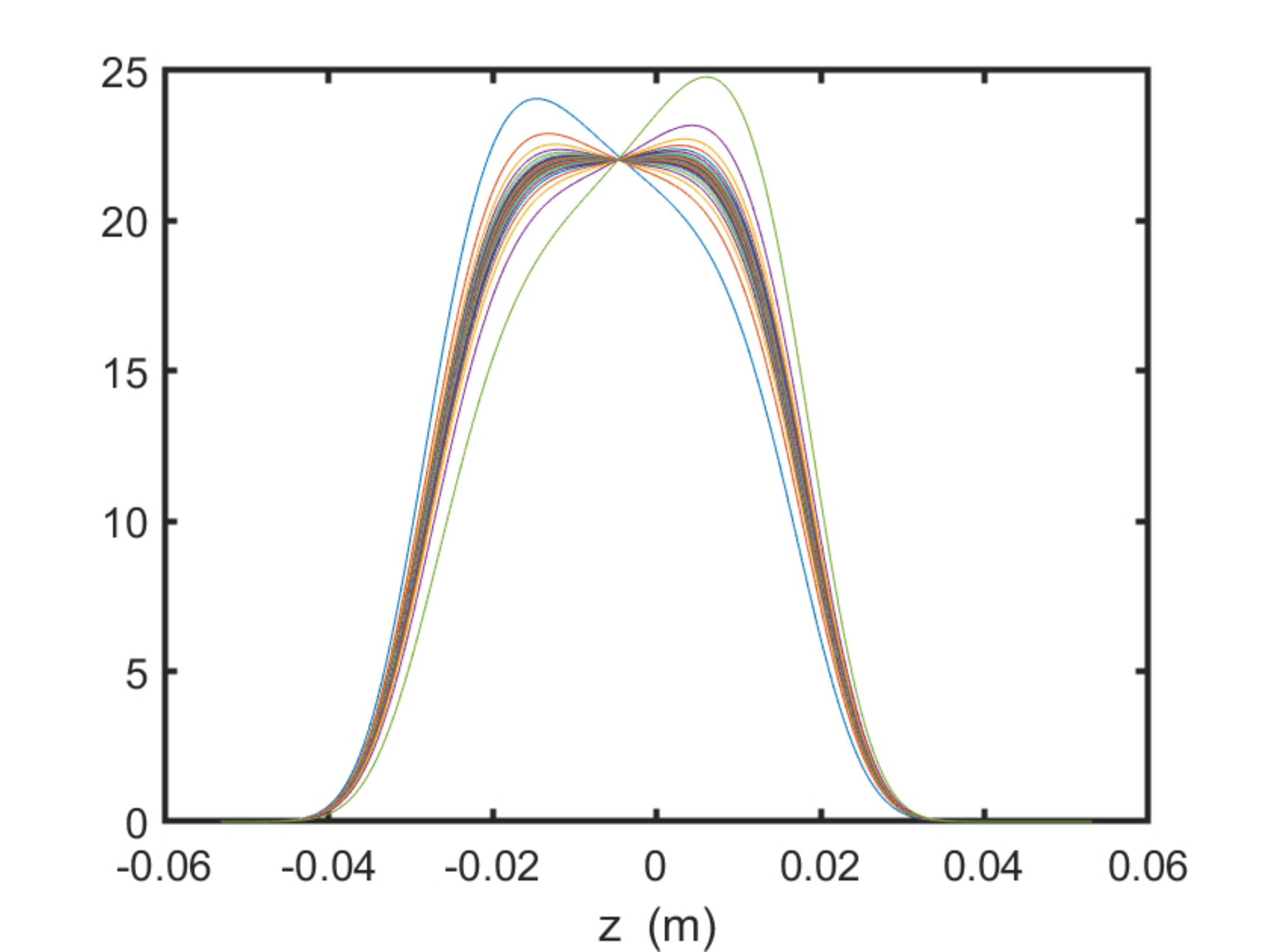}
  \caption{A train of 26 with the charge\\ distribution of Eq.(\ref{powerlaw}),  $b=1.7$\ ,  $n_g=13$.}
   \label{fig: fig14}
   \end{minipage}
  \begin{minipage} [b]{.49\linewidth}
   \includegraphics[width=\linewidth]{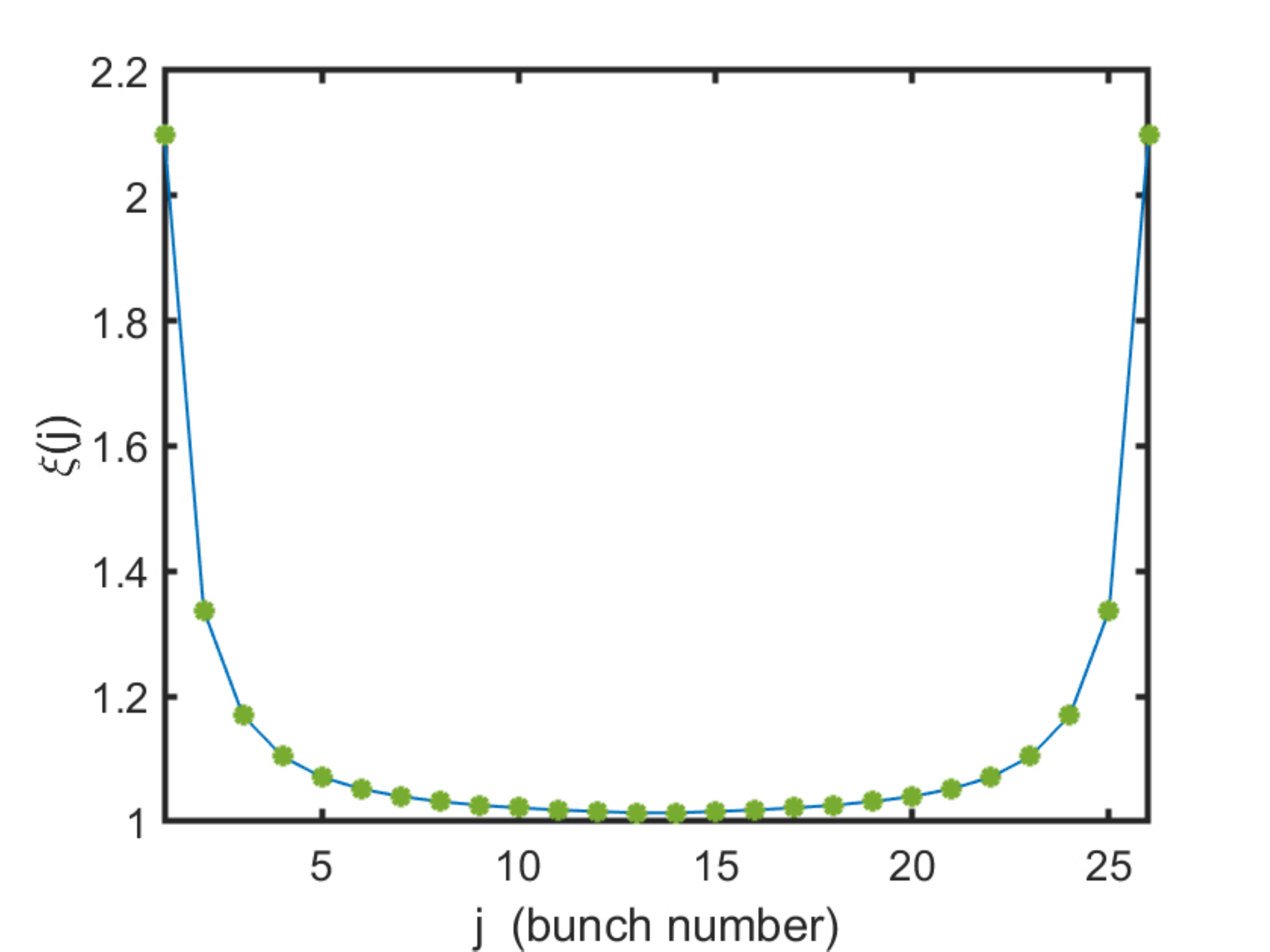}
  \caption{The bunch population factor $\xi(j)$ of Eq.(\ref{powerlaw}) for $b=1.7$ and $n_g=13$.}
   \label{fig: fig15}
\end{minipage}
\end{figure}
  If bunches at or near the ends of a train are given greater charge, enough to equal the missing charge due to the gap, the inner bunches may feel less perturbation from the gap. This idea was advanced by Byrd {\it et al.} \cite{byrd2} in 2002, and re-invented at the Argonne APS in 2017 \cite{borland}. These enhanced bunches have come to be called {\it guard bunches}. We first modify the set-up of the previous sub-section by
  increasing the charge in the first and last bunches of each of the 11 sub-trains by a factor of 3 (for example so that $\xi(1)=\xi(25)=3$), thus trying to compensate for 4 missing bunches. (Of course, it is no longer necessary to increase the total charge to compare with the complete fill.) Remarkably, this causes all of the inner bunches to be identical to graphical accuracy, with a perfect flat top as in the complete fill. Only the guard bunches differ from this pattern, as is seen in Fig.\ref{fig: fig10}.

Such highly intense guard bunches could suffer a microwave instability or have a reduced lifetime, or be undesirable for their impact on the synchrotron light pattern. It therefore becomes interesting to distribute the guard charge over several bunches. As an example we take 4 guard bunches at the beginning of the train and 4 at the end, each with 50$\%$ more charge than the inner bunches ($\xi=1.5$). As is shown in Figures \ref{fig: fig11}-\ref{fig: fig13}, the inner bunches again are flat topped, while there is a gradual transition in the guard sequence from the end bunch form to the inner.

  Rather than a uniform distribution of charge in the guard segment, one could try some kind of taper, for instance a power
  law with arbitrary exponent,
  \be
  \xi(j)=1+aj^{-b}\ ,\ j=1,2,\cdots n_g\ ,\quad \xi(j)=1+a(n_t+1-j)^{-b}\ , \quad j=n_t, n_t-1,\cdots, n_t-n_g+1\  , \label{powerlaw}
  \ee
  for a train of $n_t$ bunches with $n_g$ guard bunches at either end.  We might try a sharply peaked distribution with large $b$ in order to
  imitate the case of a single guard bunch but with less peak charge. For instance, putting $b=1.7$ and $n_g=13$ in a train with $n_t=26$
  we get the result of Figure \ref{fig: fig14}. The bunch population $\xi(j)$ of (\ref{powerlaw}) is plotted in Fig.\ref{fig: fig15}. With 30\% less charge in the end bunches we get a pattern very similar to that of a single guard bunch, in that most of the interior bunches are close to
  flat topped, and have markedly less charge than the four guard bunches with uniform population as considered above ( Figures \ref{fig: fig11} - \ref{fig: fig13}).

\subsection{Comparison to a macro-particle simulation \label{subsection:comparison}}

We have applied the code {\bf elegant} \cite{elegant} to make a macro-particle simulation for comparison to results of the present method.
This was part of an exploration of parameter space, and the parameters are different from those of Table I in the following choices:
\bea
 &&\alpha=2.07\cdot 10^{-4}\ ,\quad \sigma_\delta=1.14\cdot 10^{-3}\ ,\quad \sigma_{z0}=4.43 ~{\rm mm}\ ,\quad U_0=3.29\cdot 10^5~ {\rm eV}\nonumber\\
 &&R_s=10^6 ~\Omega\ ,\quad Q=1.67\cdot 10^4\ ,\quad \delta f=2.27\cdot 10^5 ~{\rm Hz}\ .
\eea
Also, in (\ref{firstdenb}) we take $eV\sin\phi_0=(9/8)U_0$, following Eq.(B10) in \cite{marco1}. The simulation used
the cavity wake field description provided by {\bf elegant}, and was done with 10000 macro-particles per bunch. The fill pattern is
that of Section \ref{subsection:distributed}, with 284 bunches and distributed gaps of 4 buckets each.

The agreement is good enough both to provide a check on our semi-analytic scheme and to affirm the viability of a macro-particle simulation.
\begin{figure}[htb]
   \centering
   \begin{minipage} [b]{.49\linewidth}
   \includegraphics[width=\linewidth]{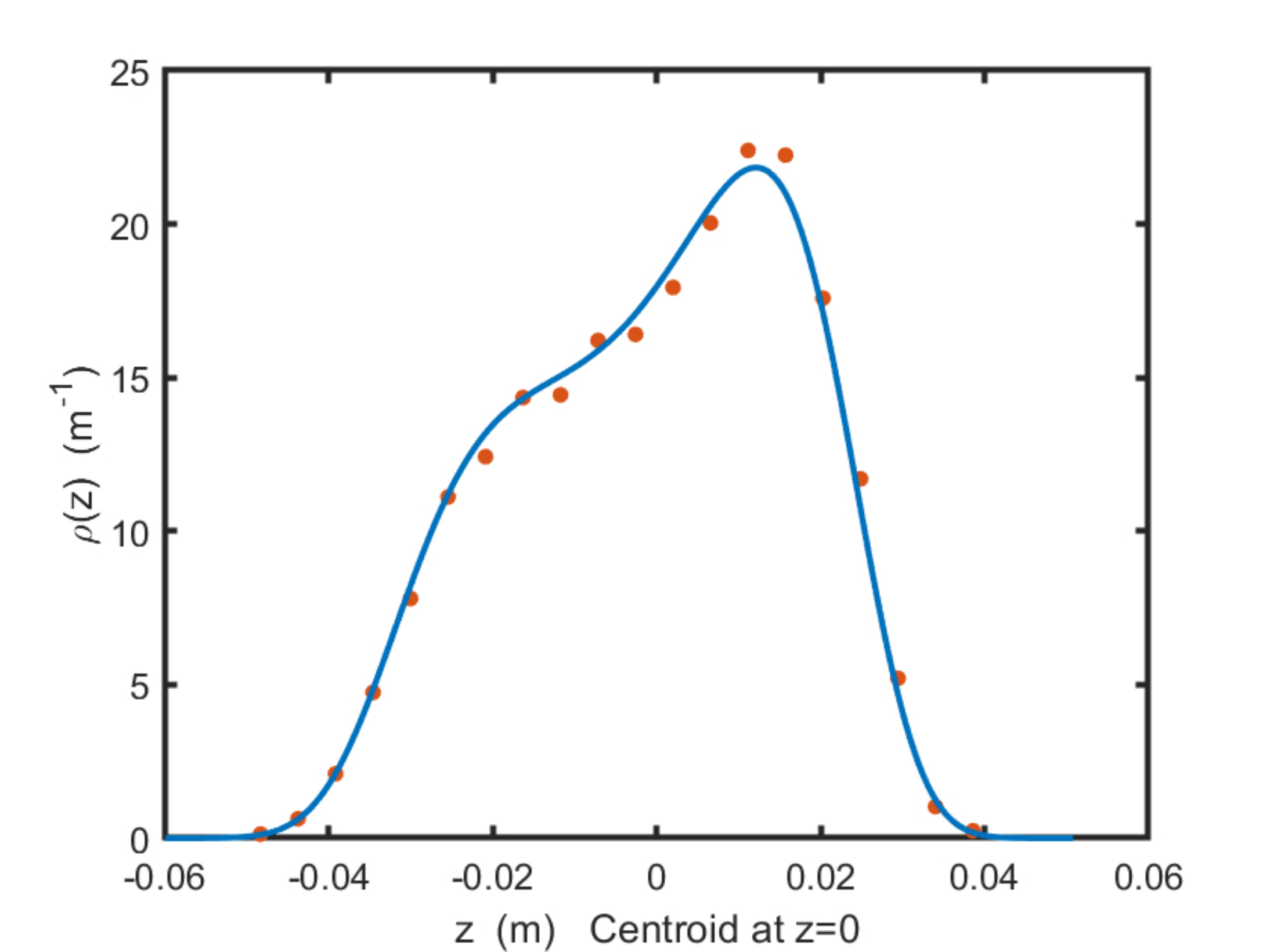}
  \caption{$\rho(z)$ of the first bunch in a train\\ of 26, by macro-particle simulation (red)\\ and Ha\" issinski solution (blue)}
   \label{fig: fig16}
   \end{minipage}
   \begin{minipage} [b]{.49\linewidth}
   \includegraphics[width=\linewidth]{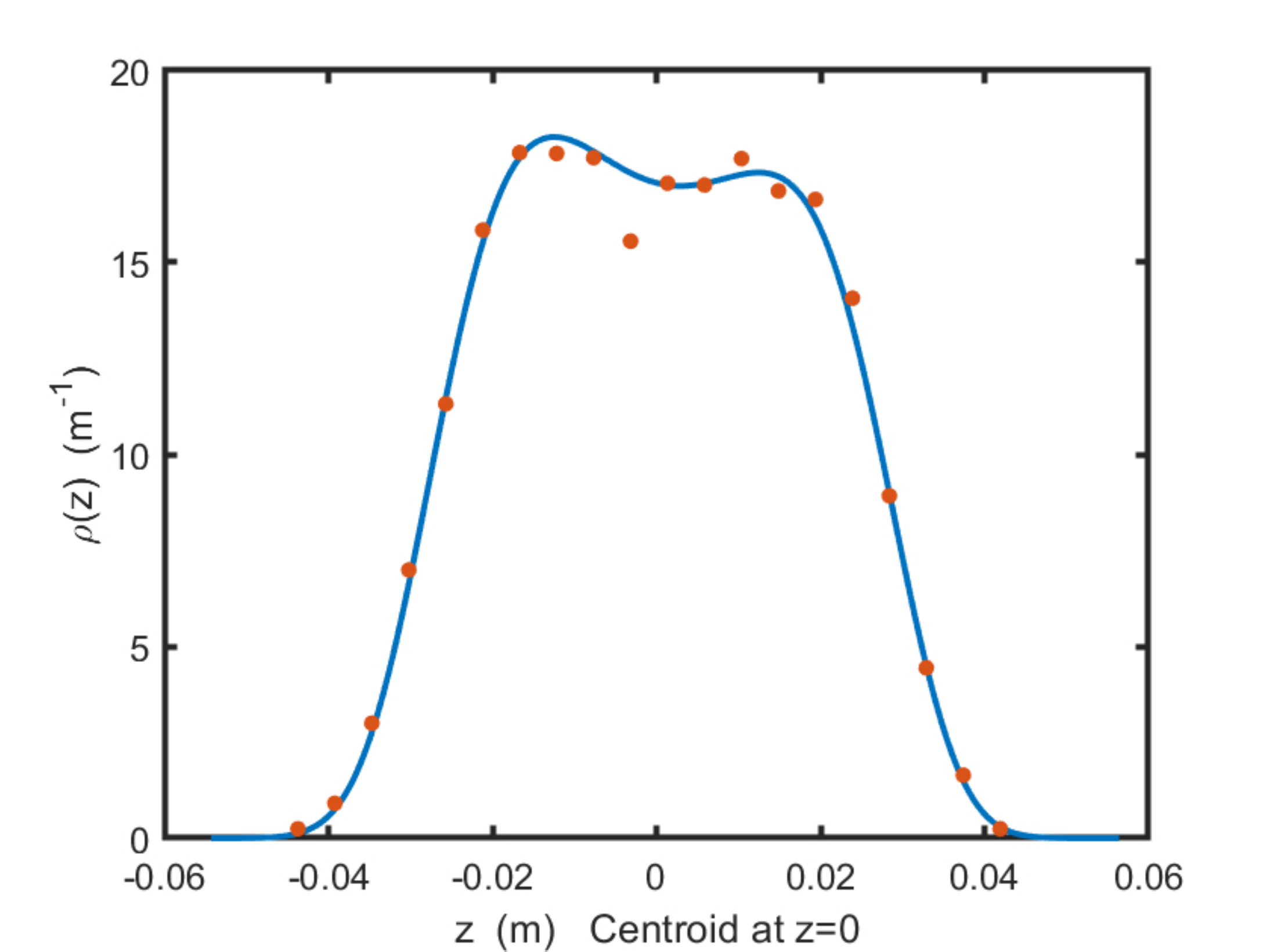}
   \caption{$\rho(z)$ of the last bunch in  a train\\ of 26, by macro-particle simulation (red)\\ and Ha\" issinski solution (blue)}
   \label{fig: fig17}
   \end{minipage}
\end{figure}
\begin{figure}[htb]
   \centering
   \includegraphics[width=.7\linewidth]{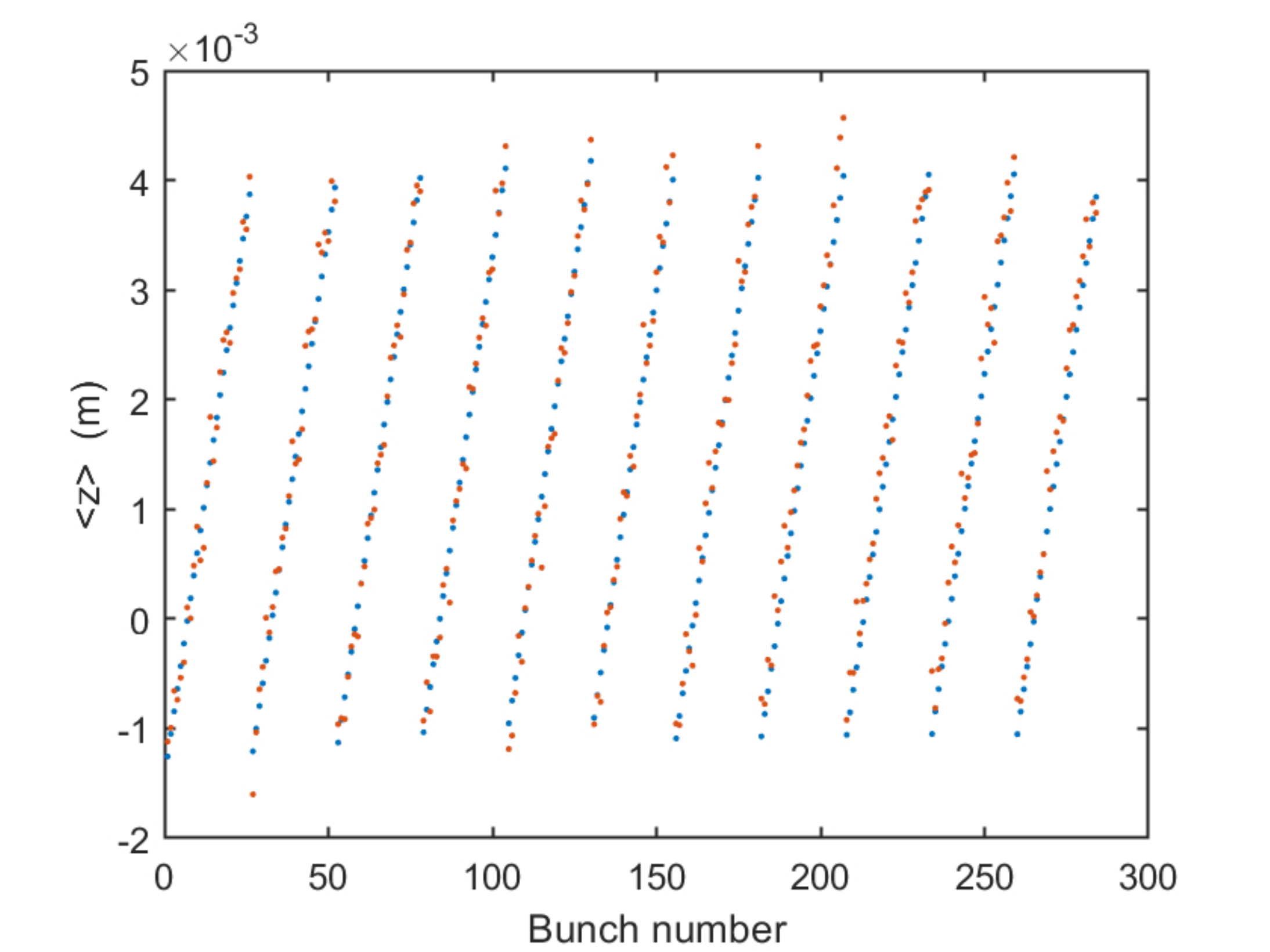}
  \caption{Centroids of a train with distributed gaps, by macro-particle simulation (red) and Ha\" issinski solution (blue)}
   \label{fig: fig18}
\end{figure}
\section{Increase in the Touschek lifetime \label{section:touschek}}
\begin{figure}[htb]
   \centering
   \includegraphics[width=.6\linewidth]{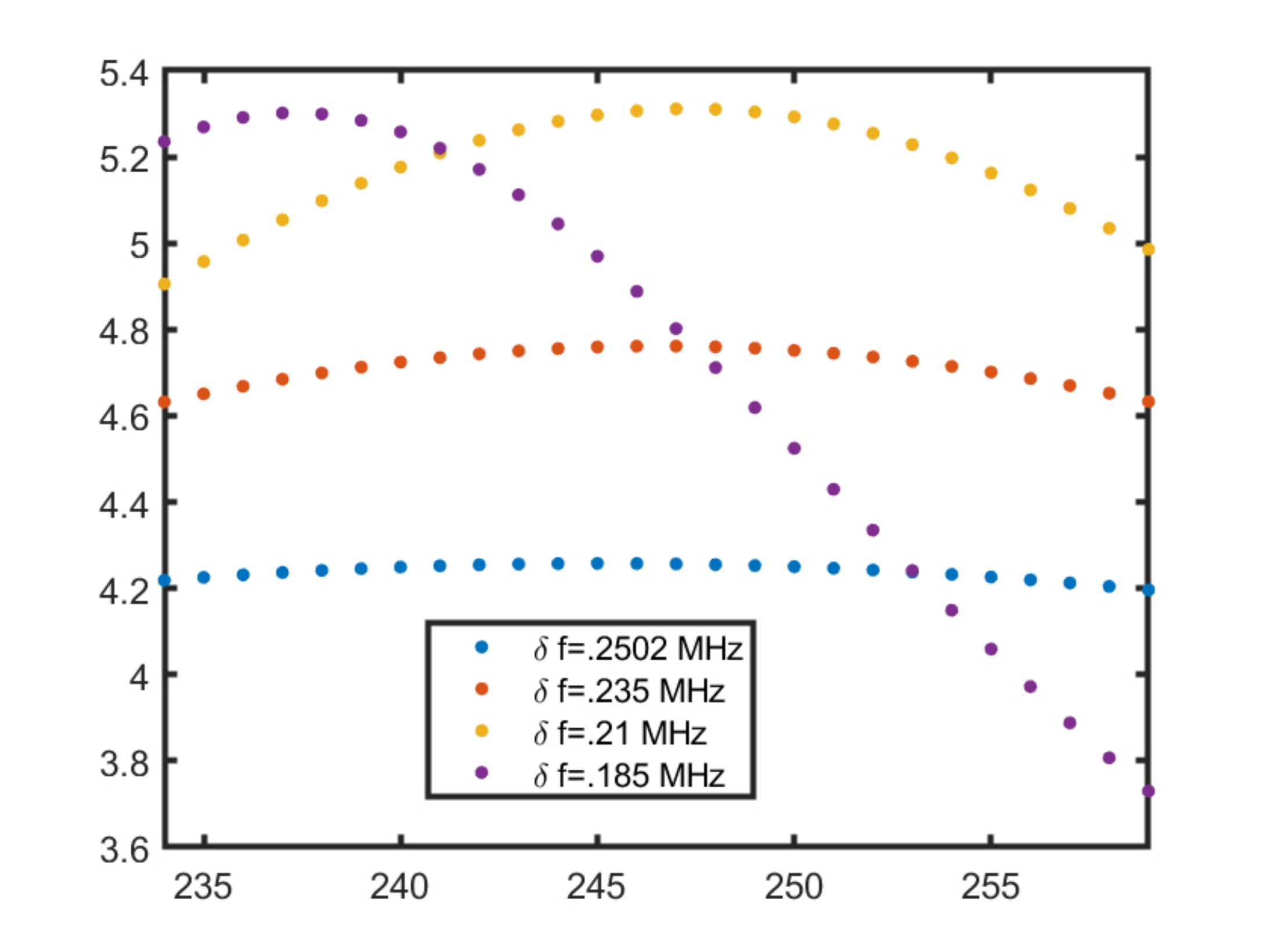}
  \caption{Ratio of Touschek lifetime to that without HHC, with decreasing detuning.
  Distributed gaps with no guard bunches.}
   \label{fig: fig19}
\end{figure}
Following Refs. \cite{byrd1} and \cite{carmignani} we note that in Piwinski's analysis \cite{piwinski} the Touschek lifetime is inversely proportional to the integral of the square of the longitudinal charge density.  Under bunch lengthening the lifetime should then increase by
a factor
\be
R=\frac{\tau}{\tau_0}\approx \frac{\int \rho^2_0(z)dz}{\int\rho^2(z)dz}\ ,                  \label{RT}
\ee
where $\rho$ and $\rho_0$ are the charge densities with and without the action of the harmonic cavity. The approximation
in (\ref{RT}) consists of neglecting the ratio of momentum acceptances for the two cases, which Byrd and Georgsson \cite{byrd1} judge to be of order 1. Taking $\rho_0$ to be a Gaussian with the nominal (zero current) bunch length, we compute $R$ by (\ref{RT}) for the case of distributed gaps, as in Section \ref{subsection:distributed}, without guard bunches.

Figure \ref{fig: fig19} shows the variation of $R$ along a train of 26,
for various choices of the cavity detuning to demonstrate the effect of over-stretching. At the nominal detuning of $\delta f=$250.2~kHz there is an almost
uniform increase of lifetime along the train, by a factor of 4.2. This occurs in spite of the substantial variation in the bunch form, suggesting
that a flat topped distribution is not a primary requirement as far as lifetime is concerned.

With smaller detuning and consequent over-stretching we get a further increase in lifetime, but with increasing variation along the train. At the smallest detuning of Fig.\ref{fig: fig19}, $\delta f=~$ 185 kHz, any advantage of over-stretching is gone, since for half of the train the
lifetime is smaller than for $\delta f=$ 235 kHz.
\section{Conclusions and outlook  \label{section:conclusions}}

We have described an effective scheme to determine the equilibrium state of an arbitrary bunch train, subject to the wake field
from a passive harmonic cavity in its fundamental mode. The calculation proceeds by an iterative method with extremely robust convergence.
The computation time is negligible, and the results agree with macro-particle simulations, which are much heavier calculations and  also much more noisy.

The quick computation allows a convenient exploration of parameter space, and in particular an examination of schemes to counter
the bad effects of gaps in the bunch train. We have seen that by distributing the empty buckets around the ring as much as possible the bunch lengthening and centroid displacement can be made comparable to those of the complete fill. Although there is then considerable deviation from
 the flat topped distribution achieved in the complete fill, that turns out not to harm the Touschek lifetime. Also, most of the bunches can be given a flat topped form by invoking guard bunches adjacent to the gaps.

 We have adopted a minimal physical model, with the only induced voltage (wake field) coming from a single resonant mode of the harmonic cavity. With  this we could illustrate the power of a new technique in the simplest way. The next step toward a realistic model should be to include the induced voltage from the main accelerating cavity (beam loading). Since our formalism allows any number of resonators, this is a straightforward extension. In fact, we have revised the code to include the main cavity, and have found that the iterative
 solution works as well as before, with only a factor of two increase in CPU time.

 Another refinement that could be significant is to include the effect of the usual short range wake fields from vacuum chamber corrugations. The magnitude of the effect on bunch lengthening can probably be judged by invoking a broad band resonator model of the
 machine impedance, which is normally applied with a Q of order 1. We have shown how to accommodate a low Q in our formalism, by retaining
  exponential factors that could have been set to 1 in the present high-Q calculations.  We have not presented the full equations for low Q,
  but those follow after replacing (\ref{W}) by the well known formula for a broad band resonator \cite{chao}, and proceeding with nearly the same steps as before. Our iterative determination of the diagonal term $v^d$ of Section \ref{section:induced} might have to be revised.

  One could also include higher order modes of cavities, and whispering gallery modes describing  coherent synchrotron radiation \cite{whispering}.

  Besides improving the physical model of the equilibrium state, an urgent matter is to study the stability of the equilibrium. This
  can of course be done by macro-particle simulations, but we would like to appeal as much  as possible to direct solution of the
  Vlasov-Fokker-Planck equation by the method of local characteristics, which proceeds with very low numerical noise \cite{senigallia}. This can
  be done easily for the case of a complete fill, with only one phase space distribution to contend with.  The present study also suggests
   possible reduced models of trains with gaps, in which identity of some of the bunches would be enforced in one way or another.
  Our technique of exploiting geometric sums can help to simplify expressions for the induced voltage.

  A special point of interest is the effect of over-stretching on thresholds of instability. We have seen, without accounting for stability, that over-stretching can give an additional increase in the Touschek lifetime.

 \section{Acknowledgments \label{acknow}}
 We thank several colleagues for advice and references to the literature: Karl Bane, Gabriele Bassi, Teresia Olsson, Boaz Nash, and Michael Borland.  For many years R. Warnock has enjoyed an affiliation with Lawrence Berkeley National Laboratory as Guest Senior Scientist.
Our work was supported in part by the U. S. Department of Energy, Contract Nos. DE-AC03-76SF00515
 and DE-AC02-05CH11231.

\appendix
\section{Second equation of motion } \label{section:appA}
To derive the second equation of motion, note that the azimuthal location of a particle with revolution frequency
$\omega_0+\Delta\omega$ is
\be
\theta=(\omega_0+\Delta\omega)t\ ,\qquad \theta=2\pi s/C\ .  \label{theta}
\ee
If the cavity is at $\theta=0$, the $n$-th passage of the cavity occurs at time $t_n$ such that
\be
2\pi n=(\omega_0+\Delta\omega)t_n\ .              \label{tn}
\ee
At that time the cavity phase is
\be
 \phi_n=\omega_1t_n=\omega_1\frac{2\pi n}{\omega_0+\Delta\omega}\approx 2\pi n\frac{\omega_1}{\omega_0}\big(1-\frac{\Delta\omega}{\omega_0}\big)=
 2\pi n h\big(1-\frac{\Delta\omega}{\omega_0}\big)\ . \label{phidmg}
\ee

The term $2\pi nh$ on the right hand side can be dropped, since it does not affect the
applied voltage $V_1\sin\phi_n$, nor the induced voltage $V_r((\phi_n-\phi_0)/k_1-m_j\lambda_1)$.
Indeed, under substitution of that term, the argument $\phi_n/k_1=\phi_n\lambda_1/2\pi$ takes on the value $nh\lambda_1=nC$. Since
$V_r$ is periodic with period $C$, it  is not changed by the presence of the term $2\pi nh$ in $\phi_n$.

With the definition (\ref{deldefs}) we then have
\be
(\Delta\phi)_{n+1}-(\Delta\phi)_n=-2\pi h\frac{\Delta\omega}{\omega_0}\ .
\ee
For highly relativistic particles above transition, the momentum compaction factor $\alpha$ can be written as
\be
\alpha=-\frac{E_0}{\omega_0}\frac{\Delta\omega}{\Delta E}\ .   \label{alphadef}
\ee
Hence with $\delta=(E_n-E_0)/E_0$ we have
\be
\frac{1}{T_0}\big[(\Delta\phi)_{n+1}-(\Delta\phi)_n\big]= 2\pi\frac{h}{T_0}\alpha\delta=\frac{2\pi}{\lambda_1}\frac{h\lambda_1}{T_0}\alpha\delta=k_1c\alpha\delta\ .
\ee
Passing to the corresponding differential equation, we obtain (\ref{secondde}).
\section{Perturbed synchronous phase \label{section:appB}}
To determine the perturbed synchronous phase   we put $k_1z_i+\phi_0=\phi_{0i}$ in (\ref{firstdenb}).  We then see that the synchronous
phase $\phi_{0i}$ for the $i$-th bunch is defined by
\be
eV_1\sin\phi_{0i}+eV_r\bigg(\frac{\phi_{0i}-\phi_0}{k_1} -m_i\lambda_1\bigg)=U_0\ ,\quad
\cos\phi_{0i}=-(1-\sin^2\phi_{0i})^{1/2}\ .  \label{newsynch}
\ee
It is the phase at which the force is zero, at the center of the distorted potential well.

In (\ref{newsynch}) we have a nonlinear equation to solve for $\phi_{0i}$.
If the equation (\ref{newsynch}) can be solved, one can work out the dynamics for a new variable $\tilde z_i$ defined by
\be
k_1\tilde z_i=\phi_i-\phi_{0i}\ ,
\ee
where $\phi_i$ is the dynamical phase of the applied voltage when the particle arrives. That is, the applied voltage takes the form $V_1\sin(k_1\tilde z_i+\phi_{0i})$, and $\tilde z_i$ is zero at the minimum of the distorted potential well. 

The scheme now involves a two-part iteration. In an iterate of Part 1 the synchronous phases $\phi_{0i}$ are determined by solving (\ref{newsynch}) with a given function $V_r$. In a succeeding iterate of Part 2, those $\phi_{0i}$ are used to calculate the charge densities and thus to form a new value of $V_r$,  by the algorithm described in Section \ref{section:newton}.

We programmed this scheme and found that it converges at moderate current, but runs into difficulty near the design current, because at that current  we are getting close to the situation in which (\ref{newsynch}) does not have a unique solution, owing to the advent of a doubly peaked charge density. At moderate current the results agree quite precisely with those from the simpler scheme based on the current-independent $\phi_0$ and the original variables $z_i$.

Since the simpler scheme works at any current up to the design current and even far beyond, we have applied it for all further work. It is not necessary to base the coordinate system on the synchronous phases, but they can be found {\it a posteriori} as the location of the minima of the distorted potential wells computed using the $z_i$ as coordinates.
\section{Reduction to the case with all buckets filled} \label{section:appC}
Here our task is to reduce our general formula for the induced voltage to its form when all r.f. buckets are filled. We shall
find that the resulting expression agrees to a close approximation with a formula well known in the literature.  Thus the following slightly complicated
 calculation   serves as a good check on the preceding work.

In the equilibrium state
the bunches will all have the same charge density $\rho(z)$. We adapt the methods of Sections \ref{section:primary} and \ref{section:effective}.
The total charge density will be
\be
\rho_{\rm tot}(z)=\sum_{p=-\infty}^\infty \sum_{j=1}^h \rho(z+(j-1)\lambda_1+pC)\ . \label{rhototeq}
\ee
By translating the integration variable we get the induced voltage as
\bea
&&V_r(z)=-eN\int W(z-z\pr)\sum_{p=-\infty}^\infty \sum_{j=1}^h \rho(z\pr+(j-1)\lambda_1+pC)dz\pr =\nonumber\\
&&-eN\int_{-\Sigma}^\Sigma\bigg[\sum_{j=1}^h\sum_{p=-\infty}^\infty W(z-z\pr+(j-1)\lambda_1+pC)\bigg]\rho(z\pr)dz\pr\ . \label{varchgd}
\eea
In the notation of (\ref{wrho})  the sum over $p$ in this expression is
\be
\mathcal{W}(z-z\pr+(j-1)\lambda_1)\ ,                   \label{calw}
\ee
which can be calculated  from the formula (\ref{weff}). To apply the formula we first show that
\be
\chi(z-z\pr+(j-1)\lambda_1)=z-z\pr+(j-1)\lambda_1\ ,              \label{chival}
\ee
to an excellent approximation. Since $0\le j-1\le h-1$ and $|z|,|z\pr|<\Sigma$, we have
\be
-\frac{1}{C}(2\Sigma+C-\lambda_1)<\frac{1}{C}(z-z\pr+(j-1)\lambda_1)< \frac{2\Sigma}{C}\ .   \label{ineq}
\ee
From this we can evaluate the ceiling function that appears in the definition (\ref{xi}) of $\chi$. At the
lower and upper bounds of its argument from (\ref{ineq}) we have
\bea
 && \lceil -1+(\lambda_1-2\Sigma)/C\rceil = 0, \quad {\rm since}\quad 2\Sigma\ll\lambda_1\ ,\label{lower}\\
 && \lceil 2\Sigma/C\rceil=1\ . \label{upper}
\eea
The evaluation (\ref{upper}) occurs only for $j=1$ and only for the part of the integration where $z\pr-z>0$. Since there
are several hundred terms of similar magnitude in the sum over $j$, this case may safely be ignored. Thus only the evaluation
(\ref{lower}) occurs, which implies that (\ref{chival}) is correct.

Using (\ref{chival}) in (\ref{calw}) and (\ref{weff}) we  now have  (\ref{varchgd}) reduced to a single sum:
\bea
&&V_r(z)=-\frac{eN\omega_rR_s\eta}{Q}\int dz\pr\rho(z\pr)\sum_{j=1}^h
\exp(-\frac{k_r}{2Q}(z-z\pr+(j-1)\lambda_1)\nonumber\\
&&\hskip 2.5cm \cdot\cos(k_r(z-z\pr+(j-1)\lambda_1)+\psi)\ . \label{onesum}
\eea
Once again, we have a geometric sum. Writing the cosine in terms of exponentials, one readily shows that
\bea
&&V_r(z)=-\frac{eN\omega_rR_s\eta}{Q}\int dz\pr\rho(z\pr)
\exp(-\frac{k_r}{2Q}\theta)\big[\cos\theta~\rep\tilde\zeta+\sin\theta~\imp\tilde\zeta\big]\ ,\label{oneeval}\\
&&\theta=z-z\pr+\psi\ ,\nonumber\\
&&\tilde\zeta(k_r)=\frac{1-r^h}{1-r}\ ,\qquad r=\exp(-k_r\lambda_1(i+1/2Q))\ .\label{tildezeta}
\eea
Note that $r$ is the same as in (\ref{polar}).

In analogy to (\ref{polar}) we define real polar variables by
\be
\tilde\zeta=\tilde\eta e^{-i\tilde\psi}
\ee
Writing
\be
\cos\theta=\cos(k_rz+\psi)\cos(k_rz\pr)+ \sin(k_rz+\psi)\sin(k_rz\pr)\ ,  \label{cosexpd}
\ee
and the analogous expression for $\sin\theta$, and applying the definition (\ref{FT}) we see that
\bea
&&V_r(z)=-\frac{2\pi eN\omega_rR_s\eta\tilde\eta}{Q}\exp(-k_rz/2Q)
\big[\cos(k_rz+\psi+\tilde\psi)\rep\hat\rho-\sin(k_rz+\psi+\tilde\psi)\imp\hat\rho\big]\ .\nonumber\\ \label{etatildeeta}
\eea

Now note the following identities and definitions:
\bea
&&\eta\tilde\eta=\bigg|\frac{1}{1-r^h}\bigg|\bigg|\frac{1-r^h}{1-r}\bigg|=\bigg|\frac{1}{1-r}\bigg|=:\kappa\ ,\nonumber\\
&& \psi+\tilde\psi=-\arg\bigg(\frac{1}{1-r^h}\bigg)-\arg\bigg(\frac{1-r^h}{1-r}\bigg)=-\arg\bigg(\frac{1}{1-r}\bigg)=\arg(1-r)
=:\varphi\ .\nonumber\\
\label{identities}
\eea
To evaluate $1-r$ notice that (for a third harmonic cavity)
\bea
&&k_r\lambda_1= 3k_1\lambda_1+\Delta k\lambda_1=6\pi+\Delta k\lambda_1\ ,\quad \Delta k=k_r-k_3\ ,\quad k_3=3k_1 \nonumber\\
&& \sin(k_r\lambda_1)\approx \Delta k\lambda_1\ ,\quad \cos(k_r\lambda_1)\approx 1\ , \label{krl1}
\eea
and
\be
1-r\approx 1-(1-k_r\lambda_1/2Q)(1-i\Delta k\lambda_1)\approx k_r\lambda_1/2Q+i\Delta k\lambda_1\ .
\ee
From this it follows that
\be
\kappa=\bigg|\frac{1}{1-r}\bigg|=\frac{f_1}{\big[(\omega_r-\omega_3)^2+(\Gamma/2)^2\big]^{1/2}}\ ,\qquad \frac{\Gamma}{2}=\frac{\omega_r}{2Q}\ ,
\ee
where $\omega=kc$ and $f_1=c/\lambda_1$ is the r.f. frequency. Thus we have Lorentzian resonant behavior, with half-width $\Gamma/2$.
Also
\be
\varphi=\arg(1-r)=\tan^{-1}\bigg[\frac{\omega_r-\omega_3}{\Gamma/2}\bigg]\ .
\ee

Since the average current is $I_{\rm avg}=eN/T_1=eNf_1$, we may write (\ref{etatildeeta}) as
\be
V_r(z)=-4\pi I_{\rm avg} R_s|\hat\rho(k_r)|\frac{\Gamma/2}{{\big[(\omega_r-\omega_3)^2+(\Gamma/2)^2\big]^{1/2}}}\cos(k_rz+\varphi+\arg(\hat\rho(k_r))
\label{vrreduced}
\ee

This is to be compared with a  formula that is often quoted in the literature; see for instance  Eq.(A7) and its derivation in Ref.\cite{marco1}.  By (A2) and (A5) of that paper and our definition (\ref{FT}),
\be
\cos\psi_{\omega_3}\approx \cos\varphi= \frac{\Gamma/2}{\big[(\omega_r-\omega_3)^2+(\Gamma/2)^2\big]^{1/2}}\ ,\qquad \hat\rho(k_r)\approx\frac{1}{2\pi}\tilde\rho(\omega_3)^*=\frac{1}{2\pi}Fe^{-i\Phi}\ .
\ee
Also, $k_rz\approx\omega_3\tau$ so that (\ref{vrreduced}) becomes (with neglect of terms higher order in $\Delta k$)
\be
V_r(z)= -2I_{\rm avg} R_sF\cos\varphi\cos(\omega_3\tau+\varphi-\Phi)\ . \label{comparemv}
\ee
in agreement with (A7) of Ref.\cite{marco1}.

\end{document}